\documentclass[apj,twocolumn]{emulateapj}
\usepackage{natbib}

\newcommand{\msun}{M$_\odot$}

\slugcomment{Submitted to ApJ.}

\shorttitle{IRS Spectral Mapping of SNR E\,0102}
\shortauthors{Sandstrom et al.}

\begin{document}


\title{Measuring Dust Production in the Small Magellanic Cloud
Core-Collapse Supernova Remnant 1E\,0102.2$-$7219}

\author{Karin M. Sandstrom\altaffilmark{1}, 
        Alberto D. Bolatto\altaffilmark{2},
	Sne\v{z}ana Stanimirovi\'c\altaffilmark{3},
	Jacco van Loon\altaffilmark{4}
	and J. D. T. Smith\altaffilmark{5}}   

\affil{$^1$Astronomy Department, 601 Campbell Hall, University of
  California, Berkeley, CA 94720, USA} 
\affil{$^2$Department of Astronomy and Laboratory for Millimeter-wave
  Astronomy, University of Maryland, College Park, MD 20742, USA}
\affil{$^3$Department of Astronomy, University of Wisconsin,
  Madison, WI 53706, USA}
\affil{$^4$Astrophysics Group, Lennard-Jones Laboratories, 
  Keele University, Staffordshire ST5 5BG, UK} 
\affil{$^5$Ritter Astrophysical Research Center, University of 
  Toledo, OH 43603, USA}	    

\email{karin@astro.berkeley.edu}


\begin{abstract}

We present mid-infrared spectral mapping observations of the
core-collapse supernova remnant 1E\,0102.2$-$7219 in the Small
Magellanic Cloud using the InfraRed Spectrograph (IRS) on the Spitzer
Space Telescope.  The remnant shows emission from fine structure
transitions of neon and oxygen as well as continuum emission from
dust.  Comparison of the mid-IR dust emission with observations at
x-ray, radio and optical wavelengths shows that the dust is associated
with the supernova ejecta and is thus newly formed in the remnant.
The spectrum of the newly formed dust is well reproduced by a model
that includes $3\times 10^{-3}$ \msun\ of amorphous carbon dust at 70
K and $2\times 10^{-5}$ \msun\ of Mg$_2$SiO$_4$ (forsterite) at 145 K.
Our observations place a lower limit on the amount of dust in the
remnant since we are not sensitive to the cold dust in the unshocked
ejecta.  We compare our results to observations of other core-collapse
supernovae and remnants, particularly Cas A where very similar
spectral mapping observations have been carried out.  We observe a
factor of $\sim 10$ less dust in E\,0102 than seen in Cas A, although
the amounts of amorphous carbon and forsterite are comparable.

\end{abstract}



\section{Introduction}
\label{intro}

Dust is a crucial component of the interstellar medium (ISM).  It
provides a site for interstellar chemistry, in particular the
formation of molecular hydrogen; it regulates thermal balance in
various phases of the ISM; and it provides shielding for dense clouds,
critically influencing the process of star-formation.  The
characteristics of interstellar dust depend on the balance between its
formation and destruction and on the processing that dust grains
endure in the ISM.  The mechanisms for dust production, their
timescales and efficiencies and the effect of metallicity on these
quantities determine the cosmic history of dust production and are
mostly not well characterized.

In particular, the relative amount of dust that core-collapse
supernovae (CCSN) contribute to the ISM is poorly constrained and a
matter of much debate.  Because of the short lifetimes of their
massive progenitor stars (a few tens of Myrs), CCSN are often cited as
an important source of dust early in the history of the Universe
\citep{morgan03,dwek07,todini01}.  Observational evidence for
substantial dust production in supernovae, however, is scant and
controversial.  

Our understanding of dust formation in the early Universe and the role
of CCSN has recently been challenged by observations of large dust
masses around high redshift quasars \citep{bertoldi03,beelen06}.
Because the redshifts of these quasars are so high and the dust masses
are so large, the dust had to be produced efficiently and on a short
timescale.  For the $z=6.42$ quasar SDSS J1148$+$5251, the age of the
Universe is only $\sim 900$ Myr and more than $10^8$ \msun\ of dust
are observed in its vicinity (although there are substantial
uncertainties associated with this dust mass, see the discussion in
\citet{omont01}).  There is some indication that dust in some of these
high redshift galaxies may follow an extinction curve unlike that in
either the Milky Way or Small Magellanic Cloud \citep{maiolino04}.
\citet{hirashita05} claim that the extinction curve of dust in the
$z=6.19$ quasar SDSS J1048$+$4637, for example, can be reproduced by
the species and size distribution of dust formed in models of the
unmixed ejecta of $\sim 10-30$ \msun\ supernovae.

The models of dust production in supernovae typically predict on the
order of $0.1-1$ \msun\ of dust formed in an average CCSN \citep[see
for instance,][]{todini01,nozawa03}. \citet{dwek07} estimate that to
produce the mass of dust observed around SDSS J1148$+$5251 requires an
average yield of $\sim 1$ \msun\ of dust per CCSN. As of yet, however,
observations of newly-formed dust in supernovae and their remnants
have shown orders of magnitude less dust than would be expected based
on these models.  If we cannot substantiate the predicted dust
production efficiency of CCSN, our explanation of the high-z quasar
observations may need to rely more heavily on the contributions of red
supergiants and massive AGB stars to dust production
\citep{vanloon05,vanloon08}.

Part of the discrepancy between expectations and observations of dust
production in CCSN and their remnants stems from the difficulty of
observing newly-formed dust.  The condensation of dust is thought to
happen between $\sim 400-800$ days after the explosion
\citep{kozasa89,todini01,nozawa03}.  While the newly-formed dust is
still warm, its near- and mid-IR emission may be detectable.
Observations of the thermal emission newly-formed from dust in SN
1987A show $\sim 10^{-4}$ \msun\ of dust produced by 775 days after
the explosion \citep{roche93,wooden93,moseley89}.  \citet{sugerman06}
claimed to detect mid-IR emission from $\sim 0.02$ \msun\ of
newly-formed dust in SN 2003gd, although this claim has been disputed
by \citet{meikle07} after their reanalysis of the Spitzer observations
that showed only $4\times 10^{-5}$ \msun\ of dust.  At these early
times, near- and mid-IR emission can also arise from IR light echoes,
as the supernova's radiation heats pre-existing circumstellar dust
\citep{bode80,dwek83,pozzo04}.  If there is a contribution by an IR
light echo, the mass of newly-formed dust inferred from the early-time
infrared emission may be greatly overestimated.  On the other hand,
clumpiness of the ejecta could lead to an underestimate of the dust
mass based on the infrared emission \citep{ercolano07}.  

As the supernova ejecta expand and cool, the newly-formed dust
becomes detectable only at far-IR and/or submillimeter wavelengths at
which time confusion with cold dust in nearby star-forming regions
becomes a major issue.  \citet{dunne03} claimed a detection of $\sim
2$ \msun\ of newly-formed dust in Cas A with 850 \micron\ SCUBA
observations.  \citet{krause04}, however, used far-IR and molecular
line observations to argue that there is at least an order of
magnitude less newly-formed dust in Cas A and that most of the
submillimeter emission arises in an intervening molecular cloud.
Similarly, submillimeter observations of Kepler's remnant were
used by \citet{morgan03} to argue for $0.1-3$ \msun\ of newly-formed
dust although these results have been recently disputed by
\citet{blair07} who find evidence for only $5\times 10^{-4}$ \msun\
using far-IR observations from Spitzer.  

Eventually, the interaction of the ejecta with the surrounding
circumstellar and/or interstellar medium (CSM/ISM) produces a reverse
shock which propagates back into the ejecta.  The reverse shock
reheats the newly-formed dust so that once again it is detectable in
the mid-IR and may destroy some of the dust via sputtering, as well.
Observations at 24 \micron\ of 1E\,0102.2$-$7219, the same SNR we
discuss here, by \citet{stanimirovic05} showed evidence for $\sim
8\times 10^{-4}$ \msun\ of dust. Recently, \citet{rho08} used IRS on
Spitzer to make a spectral map of Cas A.  They claim evidence for
$\sim 0.02$ \msun\ of newly-formed dust in the remnant.  Since only
the reverse-shocked dust should be visible in the mid-IR, only a
fraction of the newly-formed dust can be detected in this way.
Additionally, circumstellar or interstellar dust that is interacting
with the forward shock should also be detected in the mid-IR.
\citet{williams06} observed four supernova remnants in the Magellanic
Clouds with the IRAC and MIPS instruments on Spitzer and concluded
that the mid-infrared emission was from shocked circumstellar dust
rather than newly-formed dust.  Despite these limitations mid-IR
observations are very useful for studying the dust production in
supernovae because they typically have higher angular resolution,
which allows the remnant to be more easily separated from foreground
and background emission.  Mid-IR spectroscopy also has the potential
to provide some information on the composition of the newly-formed
dust in a remnant from the distinctive mid-IR spectral features of
some dust species.  In this paper we present mid-IR spectral mapping
observations of the Small Magellanic Cloud remnant 1E\,0102.2$-$7219,
which we use to learn about the mass and composition of dust produced
in CCSN.

SNR 1E\,0102.2$-$7219 (E\,0102) is a young, oxygen-rich remnant
located on the outskirts of the emission-line nebula N\,76.  It is
thought to be the product of a Type Ib/Ic or IIL/b supernova
\citep{blair00,chevalier05} due to the lack of hydrogen and helium in
the remnant's spectrum and the high abundance of oxygen and neon in
its ejecta.  E 0102 has been observed extensively at x-ray, UV,
optical, infrared and radio wavelengths.  Chandra x-ray imaging of the
remnant by \citet{gaetz00} showed a textbook remnant structure with a
blast wave radius of $\sim$ 22'' and a reverse shock radius of $\sim$
15''. \citet{amy93} observed the remnant at 6 cm using the Australia
Telescope Compact Array (ATCA) and found a shell of radio emission
with a radius of $\sim$ 20''.  This ring lies between the measured
radii of the faint rim of the blast wave and the bright ring of the
reverse shock, and is attributed to material in the forward shock.
The full-width velocity dispersion of the optical [O III] emission
from the remnant was measured to be $\sim 6000$ km/s by
\citet{tuohy83}, implying an age of $\sim 1000$ years given the
remnant's size.  \citet{hughes00} use three epochs of x-ray
observations from the \textit{Einstein}, \textit{ROSAT} and
\textit{Chandra} observatories to measure an expansion rate for the
whole remnant in agreement with the results of \citet{tuohy83}.
Recent work by \citet{finkelstein06} used two epochs of optical [O
III] observations to determine proper motions for ejecta filaments and
find an age of $2050\pm 600$ years for the remnant.  The results of
the following work do not depend sensitively on the age, so we adopt a
value of 1000 years for simplicity.

\citet{stanimirovic05} presented the first infrared detection of SNR E
0102 from the Spitzer Survey of the Small Magellanic Cloud (S$^3$MC)
observations.  S$^3$MC covered the star-forming regions of the SMC Bar
and Wing with imaging in all MIPS and IRAC bands \citep{bolatto07}.
The remnant was only clearly detected at 24 \micron.
\citet{stanimirovic05} used the 24 \micron\ detection along with the
upper limits in the other bands to argue that the dust in the remnant
has a temperature of $\sim 120$ K and they determined an upper limit
of $8\times 10^{-4}$ \msun\ of warm dust. The main uncertainties in
their analysis were the fraction of the MIPS 24 \micron\ emission
coming from the [O IV] line at 24.8 \micron\ and the temperature of
the dust.

In the following, we present Spitzer IRS spectral mapping observations
of SNR 1E\,0102.2$-$7219 obtained as part of the Spitzer Spectroscopic
Survey of the Small Magellanic Cloud (S$^4$MC).  These observations
allow us to map the emission lines and dust continuum from the remnant
and place a lower limit on the amount of newly formed dust in E 0102.
In Section~\ref{sec:data} we describe the observations and their
reduction.  In Section~\ref{sec:spec} we present the spectrum of the
remnant and discuss the spatial distribution of its various
components.  In Section~\ref{sec:dustmodel} we use a model for the
reverse-shocked dust to measure a mass of $3\times 10^{-3}$ \msun\
and we discuss the composition of the dust as revealed by its
spectrum.  Section~\ref{sec:discussion} discusses the implications of
our measurement for the understanding of CCSN dust production and
compares our results with previous work.

\section{Observations and Data Reduction}

\subsection{Spitzer Observations}\label{sec:data} 

SNR E 0102 was mapped with all LL and SL orders of IRS on the Spitzer
Space Telescope as part of the S$^4$MC project (Spitzer Spectroscopic
Survey of the Small Magellanic Cloud, GO 30491).  The
wavelength coverage of the SL and LL orders extends from 5.2 to 38.0
\micron\ with a spectral resolution between 60 and 120.  Ramp times of
14 and 30 seconds for SL and LL were chosen as a compromise between
sensitivity, spatial coverage and observation length.   

\begin{figure*}[t]
\centering
\epsscale{1.0}
\includegraphics[width=5in]{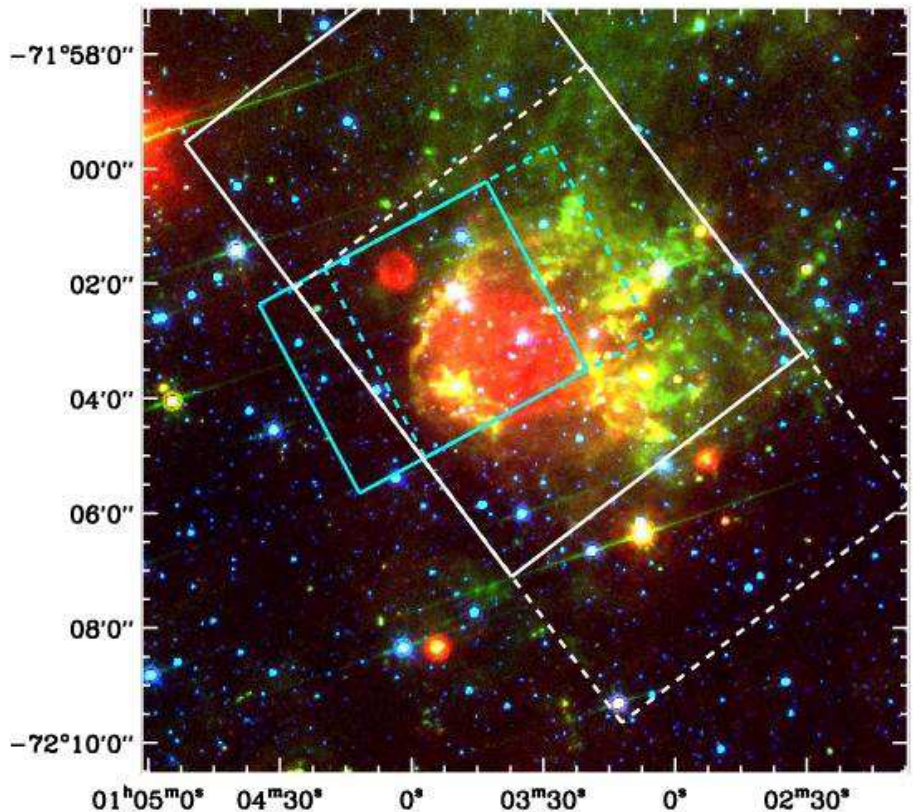}
\caption{Three-color image of the N 76 region
overlayed with the coverage of the IRS spectral mapping observations.
The S$^3$MC MIPS and IRAC maps at 24, 8.0 and 3.6 \micron\ are shown
in red, green and blue.  The white solid and dashed lines show the
coverage of the LL1 and LL2 spectral maps.  The cyan solid and dashed
lines show the coverage of the SL1 and SL2 maps. The 1E\,0102.2$-$7219
supernova remnant is located at R.A.  $01^{\rm h}04^{\rm m}02\fs1$,
Dec.  $-72^\circ01\arcmin52\farcs5$ (J2000).}
\label{fig:n76}
\end{figure*}

SNR E 0102 was covered in the spectral maps of the N\,76 region,
observed on 9 and 12 December 2006 for SL and LL, respectively. The LL
map consists of $6\times 75$ pointings covering an area of $474\times
381$ arcseconds$^2$ in both LL1 and LL2.  The SL observations
consisted of $5\times 120$ pointings covering $260\times 222$
arcseconds$^2$ for the SL1 and SL2 orders.  The roll angle was left
unconstrained for our observations. The SL and LL maps have different
orientations because of the slit positions in the focal plane of the
telescope.  Figure~\ref{fig:n76} shows a three-color image of the
region from the S$^3$MC dataset overlayed with the coverage of the
SL2, SL1, LL2 and LL1 maps.  The spectral maps are completely sampled
in both LL and SL by stepping perpendicular to the slit by one-half
slit width (5.08 and 1.85'' for LL and SL, respectively).  In the SL
orders we have stepped by a full slit width parallel to the slit
(52'') in order to increase the spatial coverage of the maps at the
expense of some redundancy.  In the LL orders we have used half-slit
width steps parallel to the slit (79'').  More redundancy in the LL
maps is necessary due to the increasingly large numbers of bad pixels
at long wavelengths.  

\subsection{Data Reduction}\label{sec:reduction}

The spectral maps were calibrated and assembled in IDL using the
Cubism package \citep{smith07}.  Cubism takes as input
pipeline-processed two-dimensional spectral images from IRS and
produces three dimensional spectral data cubes using a reprojection
algorithm which employs two steps of polygon clipping (in which the
exact geometrical overlap of input pixels, or pseudo-rectangles in the
case of the IRS spectral images, is used to calculate its weight in an
output pixel grid).  Cubism also applies an aperture loss correction
function (ALCF) and a slit loss correction function (SLCF) to adjust
the calibration which is based on point-source observations rather
than extended sources. The output pixel scales of the LL and SL cubes
are 5.08'' and 1.85'', respectively.  Images of the IRS data in this
work are shown on the IRS pixel scale rather than aligned to R.A. and
Dec to avoid unnecessary interpolation.

The Galactic and zodiacal light foregrounds were removed from the IRS
spectra using a dedicated ``off'' position.  The same position was
used for both the LL and SL modules.  The background observations were
done immediately following the mapping observations.  The ``off''
position was chosen as a place in the S$^3$MC 24 \micron\ MIPS mosaic
with no extended emission or point sources, located at R.A.  $1^{\rm
h}09^{\rm m}40\fs00$, Dec.  $-73^\circ31\arcmin30\farcs00$.  

The subtraction of the ``off'' position, in addition to removing the
Galactic and zodiacal foregrounds, mitigates the effects of rogue
pixels (i.e. time variable bad pixels) in the IRS detectors, although
it does not eliminate them.  Further bad pixel removal was done in
Cubism by examining each wavelength for striping (caused by global bad
pixels) and other artifacts.  The effect of bad pixels increases at
the longer wavelengths, degrading the data quality beyond about 35
\micron. 

\subsection{Resolution Matching and Alignment of
Cubes}\label{sec:conv}

In order to directly compare the spatial distribution of supernova
emission at each wavelength as well as properly extract a spectrum
when the extraction region is on the order of a resolution element, it
was necessary to convolve all the individual slices of the spectral
cube to the same resolution.  This was achieved by creating
convolution kernels based on theoretical point-spread functions (PSFs)
for the IRS spectrograph generated by the program sTinyTim
\footnote{\url{http://ssc.spitzer.caltech.edu/archanaly/contributed/stinytim/index.html}}.

\begin{figure*}
\centering
\epsscale{1.0}
\includegraphics[width=5in]{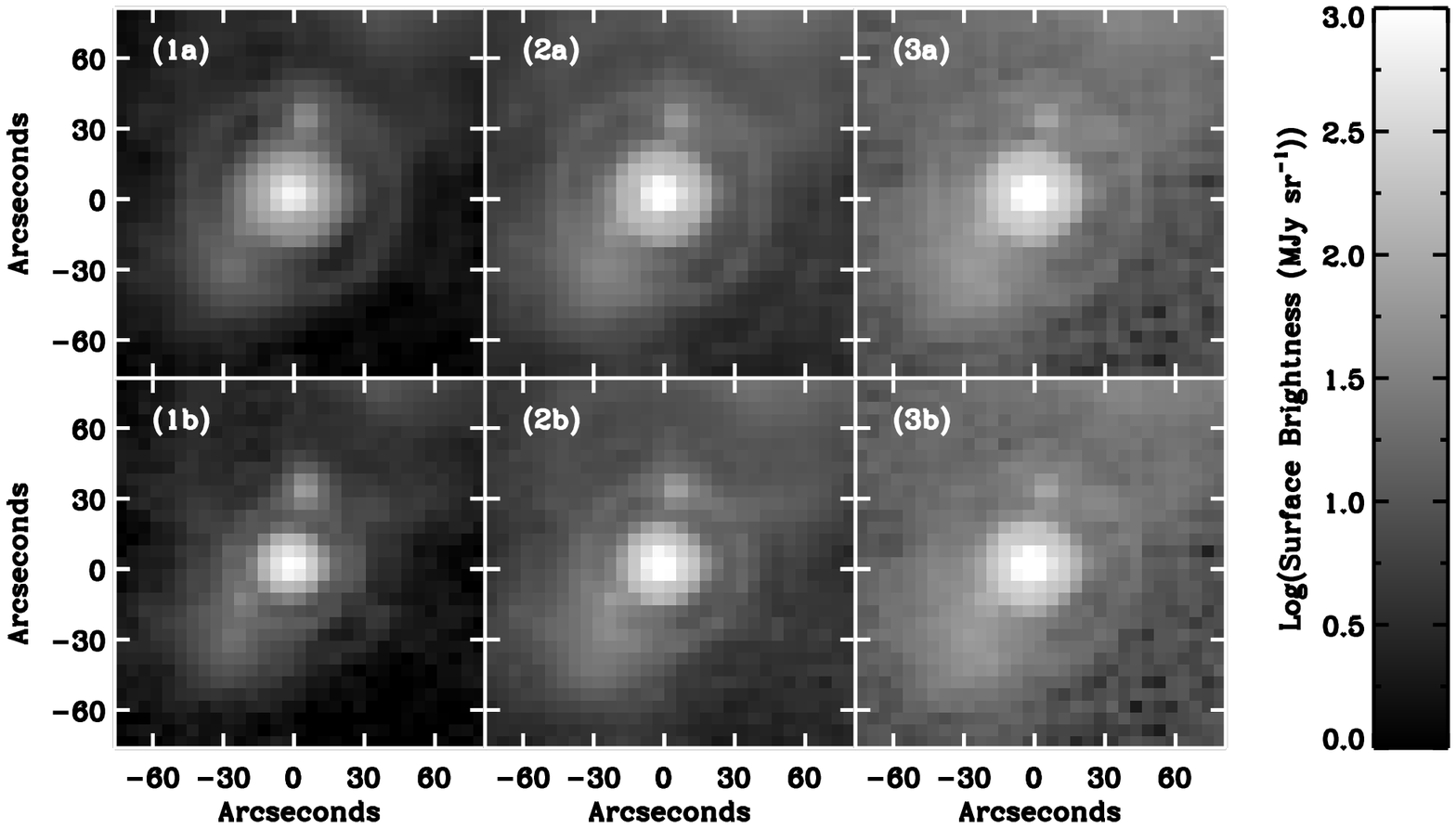}
\caption{Results of the convolution described in 
Section~\ref{sec:conv} for a bright point source in the N\,22 region.
The panels show the point source at (1) 23.0 \micron, (2) 30.0 \micron,
and (3) 36.7 \micron. The bottom panels (1b), (2b) and (3b), show the
point source before the convolution. The top panels, labeled (1a),
(2a) and (3a) show the source after convolution.  The image stretch is
logarithmic to highlight the faint diffraction rings around the point
source.  After the convolution, the diffraction rings are aligned at
all wavelengths, showing the effectiveness of our sTinyTim derived
convolution kernels.}
\label{fig:n22}
\end{figure*}

The convolution kernels $k$ were defined in the following way:
\begin{equation} 
k = \mathrm{FFT}^{-1}(R/I) ,
\end{equation} 
where $R$ is the 2-dimensional Fast Fourier Transform (FFT) of the PSF
of the longest wavelength slice of the LL1 cube and $I$ is the FFT of
the PSF of the slice in question.  The convolution kernels were masked
in Fourier space such that there was no power at spatial frequencies
higher than the Nyquist frequency of the lowest resolution kernel in
order to avoid the amplification of high-frequency noise.  Each slice
of the spectral data cube was then convolved with the appropriate
kernel.  Although the PSF of IRS is not well explored (i.e. the
predictions from sTinyTim are mostly untested) and the PSF that Cubism
recovers can be somewhat elongated in a way that varies with
wavelength, we found that the results of this technique were adequate
for our purposes.  The convolution process was tested on the cube of
the N 22 region of the SMC which contains a very bright point source
for which the first two diffraction rings are visible for essentially
every slice in the cube from SL2 through LL1.  The correspondence
between wavelength slices after our convolution is excellent as shown
in Figure~\ref{fig:n22}.  For each cube, the associated uncertainty
cube was also convolved and the errors appropriately propagated.

The LL1, LL2, SL1 and SL2 cubes were aligned using a technique based
on the polygon-clipping algorithm implemented in Cubism.  In
polygon-clipping the weight assigned to each pixel in the original
image is calculated based on the geometric overlap of that pixel with
the grid of output pixels. Aligning the cubes using polygon clipping
minimizes interpolation error and  allows for direct propagation of
the associated uncertainty cubes. The LL2, SL1 and SL2 cubes were
aligned to match the astrometry and pixel scale of the LL1 map after
resolution matching.

\subsection{Local Background Removal}

More difficult to remove than the Galactic foreground is the variable
and complex local foreground and background emission (for simplicity
we will refer to this as a background, though there are contributions
from the entire line of sight through the SMC).  E 0102 is located
near the large emission-line complex N\,76.  Emission from the
outskirts of this region must be removed from the supernova remnant
spectrum to accurately measure the dust content and emission line
flux.  

An accurate background subtraction is not possible over a large area
due to the proximity of N\,76 and the complexity of the emission in the
region.  For this reason we focus on the area near the supernova
remnant and attempt to subtract a model for the local emission.  We
first attempted to fit planar or higher order surfaces to the emission
in the region of the remnant.  The steep profile on the side of N\,76
nearest the remnant was poorly represented in these fits, resulting in
large residuals.  Polynomial fits to each row and column, as done by
\citet{stanimirovic05}, were a better representation of the local
emission but also suffered from issues with reproducing the steepness
of the N\,76 shoulder, interfering with the goal of determining the
background in the immediate vicinity of the remnant.  Rather than
trying to reproduce the local emission we argue that a better
representation of the background comes from interpolating across the
supernova remnant, which eliminates the need to reproduce the complex
emission structures with polynomials or surfaces.  

For each slice of the convolved and aligned cubes, we mask out the
immediate region of the remnant, using a mask 60'' in diameter to
exclude any emission from the remnant in our background determination.
We then linearly interpolate in the $x$ and $y$ directions across the
masked region, using two pixels on either side of the remnant in the
given row or column.  Finally, we average together the results
obtained by interpolating in the $x$ and $y$ direction.  This
technique has advantages and disadvantages.  Since we only use the
four pixels nearest the remnant, there is some random noise in the
background value.  We believe this disadvantage is far
outweighed by the difficulty in doing any kind of more complex fit
given the spatial intensity variations of the background.  In
addition, the column based interpolation also tends to remove any
residual striping in the LL cubes that results from the effects of
rogue detector pixels.

\begin{figure*}
\centering
\epsscale{1.0}
\includegraphics[width=5in]{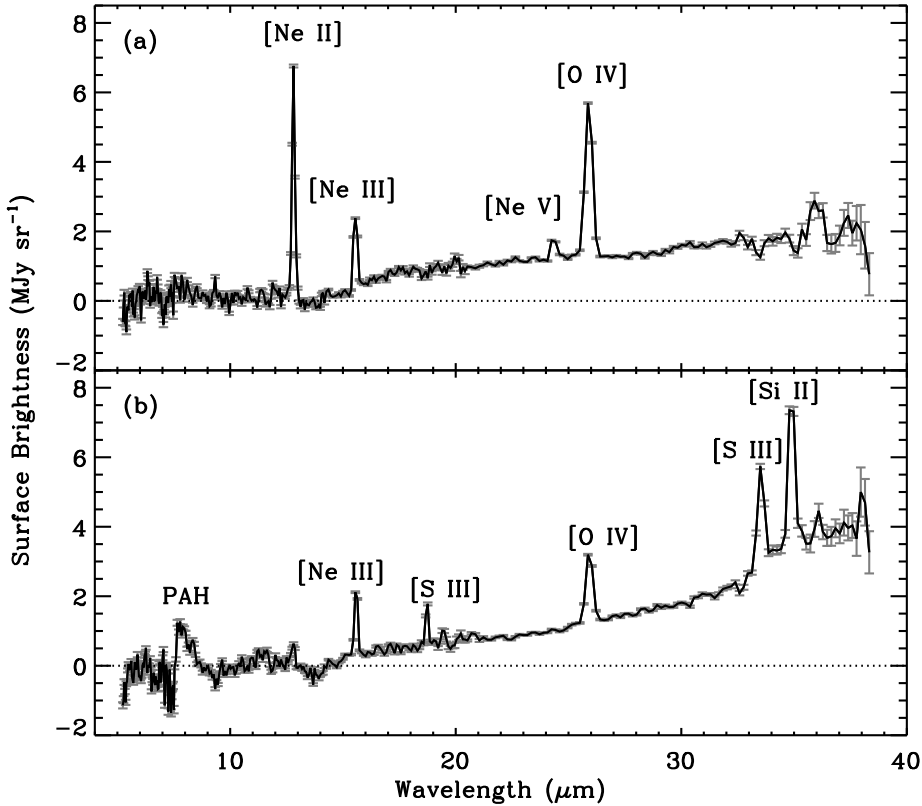}
\caption{This figure shows (a) the spectrum of SNR E 0102 
after the subtraction of the local background and (b) the local
background spectrum.  The spectrum of the supernova remnant shows dust
continuum and emission lines from neon and oxygen.  The background
spectrum shows lines from sulfur, silicon, neon and oxygen as well as
emission from polycyclic aromatic hydrocarbons around 8 \micron\ and
dust continuum.  The contributions from silicon, sulfur and PAHs are
entirely removed from the remnant's spectrum after background
subtraction.  The background spectrum is slightly negative at short
wavelengths due to oversubtraction of the Galactic and zodiacal light
foregrounds.  The background subtraction corrects this oversubtraction
in the spectrum of the remnant.} 
\label{fig:snr}
\end{figure*}

A plot of the background spectrum is shown in the bottom panel of
Figure~\ref{fig:snr}.  Subtracting the background eliminates the
contributions of [S III], [Si II] and PAHs entirely from the SNR
spectrum showing that emission from these species is not related to
the remnant.  All of these emission features are typical for the ISM
irradiated by the average interstellar radiation field (ISRF).  The
local background spectrum is slightly negative at short wavelengths.
This is most likely due to the gradient in the zodiacal and Galactic
foregrounds between the location of the dedicated ``off'' position and
the remnant. Subtracting the local background, however, corrects for
this over-subtraction in the spectrum of the remnant.  In the
continuum, background emission accounts for 40 to 60\% of the
emission.  For the emission lines, the background subtraction removes
a few percent of the integrated strength of [Ne III] and [O IV].

In general, the local emission in each wavelength slice has a very
similar spatial structure, except at the wavelengths of strong
emission lines.  In the strong emission line from [O IV] and/or [Fe
II] at 25.8 \micron, a diffuse halo is evident around the supernova
remnant. This ionized gas cannot be from material that is
collisionally interacting with the remnant since it is well outside
the extent of the forward shock.  A similar structure is seen at the
wavelengths of the other spectral lines.  The ionized material
surrounding the remnant has been observed previously in optical
studies \citep{tuohy83,blair00,finkelstein06}.  The source of the
ionization for this surrounding material has been discussed by
\citet{tuohy83}.  It is possible that the material was ionized by the
supernova shock breakout or is currently being ionized by emission
from the radiative shocks in dense knots of ejecta in E 0102.  The
surroundings of the remnant will be discussed further in
Section~\ref{sec:csm}. 

\section{Analysis}

\subsection{The Mid-IR Spectrum}\label{sec:spec}

\subsubsection{Sources of Dust Continuum Emission}\label{sec:temps}

We now consider the possible sources for dust continuum emission from
SNR E 0102: circumstellar/interstellar dust in the forward shocked
region, newly formed dust in the reverse shocked ejecta or newly
formed dust in the unshocked ejecta.  The contribution of each of
these regions to the spectrum depends on the mass, temperature and
composition of the dust present.  

The dominant heating process for dust in the shocked ejecta and
shocked CSM/ISM will be collisions with electrons
\citep{dwek81,arendt99}.  The equilibrium temperature of a dust grain
that is collisionally heated in a plasma can be found by equating the
collisional heating rate of the grain with its radiative cooling rate.
The collisional heating rate for a dust grain with size $a$ by
particles with mass $m$, density $n$ and temperature $T$ is given by 
\citet{dwek87}:
\begin{equation}
H(a,T,n) = (32/\pi m)^{1/2} \pi a^2 n(kT)^{3/2} h(a,T) .
\end{equation}
Here $h(a,T)$ is a dimensionless parameter which describes the grain
``heating efficiency'' and depends on the fractional amount of the
incident energy of the particle that is deposited in the grain by the
collision.  \citet{dwek87} develop an approximation for $h(a,T)$ based
on a comparison of the effective thickness of the grain $4a/3$ and the
experimentally measured stopping distance of an electron or ion in a
solid. We use their formulae in this analysis.  The radiative cooling
rate of the grain is given by:
\begin{equation}
L_{gr}(a, T_d) = 4\pi a^2 \sigma T_{d}^{4} <Q(a,T_d)> ,
\end{equation}
where $<Q(a,T_d)>$ is the Planck-averaged value of the absorption
coefficient for a dust grain of the given size and temperature.
Equating these heating and cooling rates yields the equilibrium
temperature of the grain.  

\begin{figure}
\centering
\epsscale{1.0}
\includegraphics[width=3in]{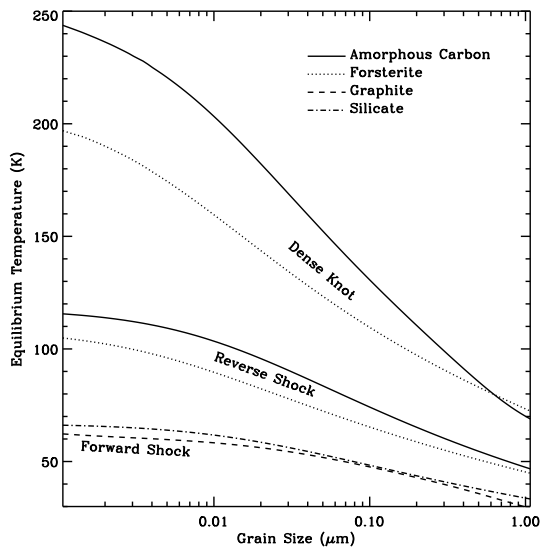}
\caption{The equilibrium temperature of collisionally heated grains
of different compositions in the forward and reverse shocked x-ray
emitting plasma and the hot, post-shock region of a dense knot. We
plot only amorphous carbon and forsterite in the reverse shocked ejecta
based on the results of our dust model fits in Section~\ref{sec:fit}.
We assume that dust grains in the shocked CSM/ISM will have the
optical properties characteristic of interstellar graphite and
silicate from \citet{laor93}.}
\label{fig:teq}
\end{figure}

In the remnant, shocked CSM/ISM dust would be located in a shell with
an outer radius of 22'' and an inner radius at the contact
discontinuity, the exact position of which has not been determined.
X-ray spectroscopy of the blast wave region shows post-shock material
with abundances consistent with the SMC, electron temperatures around
$1$ keV, and ionization timescales $\tau = n_e t \sim 3\times 10^{10}$
s cm$^{-3}$ \citep{hayashi94,hughes00,sasaki06}.  Assuming a timescale
of $\sim 1000$ years (the age of the remnant) gives electron densities
in the forward shock region of $n_e \sim 1$ cm$^{-3}$ .  Collisionally
heated dust grains in this region would have equilibrium temperatures
of between $30$ and $70$ K for silicate and carbonaceous grains with
sizes between $0.001$ and $1.0$ \micron\ \citep[assuming the optical
constants for interstellar graphite and silicate grains
from][]{laor93}.  Figure~\ref{fig:teq} shows the equilibrium
temperature of dust grains in the forward shocked plasma as a function
of grain size. The collision rates in this plasma, as well as in the
reverse shock, are low enough that the grains are not heated above
their sublimation temperature and destruction of the grains via
evaporation.

Shocked ejecta dust would be located in a shell with an outer radius
at the contact discontinuity and an inner radius at 15''.  The heating
mechanism of the dust in the shocked ejecta depends on its density
structure.  When the reverse shock encounters more diffuse parts of
the ejecta its passage creates hot x-ray emitting plasma.  X-ray
spectroscopy of the shocked ejecta shows a plasma composed of oxygen,
neon, magnesium and silicon with electron temperatures around 0.4 keV
($5\times 10^6$ K) and ionization timescales of $n_e t \sim 10^{12}$ s
cm$^{-3}$.  Assuming $t\sim 1000$ years, as before, gives densities in
the x-ray emitting reverse shocked ejecta of $\sim 20$ cm$^{-3}$.
Figure~\ref{fig:teq} also shows the equilibrium temperature of
collisionally heated amorphous carbon and forsterite (Mg$_2$SiO$_4$)
grains as a function of size given these plasma parameters, using
optical constants from \citet{rouleau91} for amorphous carbon and
\citet{jaeger03} for forsterite and assuming the heating efficiencies
for carbonaceous and silicate dust from \citet{dwek87}.  

The interaction of the reverse shock with material in dense clumps of
ejecta drives radiative shocks that produce optical and infrared line
emission, discussed further in Section~\ref{sec:lines}, and heat the
dust grains by a combination of radiative and collisional processes.
\citet{arendt99} compared the collisional and radiative heating for
dust in the hot, post-shock region of a knot in Cas A assuming the
density, temperature and cooling function predicted by the models of
\citet{sutherland95} for shock velocities between $150-200$ km
s$^{-1}$---the same models which are found to best reproduce the
optical and UV emission from E 0102.  They found that the collisional
heating exceeds radiative heating under these conditions by factors of
a few hundred. Thus, for dust in the immediate post-shock part of
dense knots, we can ignore radiative heating.  Figure~\ref{fig:teq}
shows the equilibrium temperatures for collisionally heated dust
grains in this region, using $T_e = 10^{6.64}$ K, $n_e = 400$
cm$^{-3}$ and the gas cooling function $\Gamma = 10^{-17.5}$ ergs
cm$^{3}$ s$^{-1}$ from \citet{sutherland95}.  On the other hand, in
the cooled post-shock gas of the dense knots and the in cold pre-shock
gas, the dust is likely heated by the radiation field generated by the
shock front.  \citet{bouchet06} calculate the radiative heating for
dust in similar conditions in SN 1987A and find that it can reach
temperatures of $\sim 125$ K.  Given the angular resolution of our
observations we will not be able to distinguish regions where dust in
knots is collisionally or radiatively heated, but we note that the
equilibrium dust temperatures in these circumstances are all between
$\sim$ $100-200$ K.

If the dust grains in these various collisionally heated regions are
small enough, a collision with a single electron may be sufficient to
raise the grain temperature well above the equilibrium value for a
short time.  After this temperature spike the grain can cool to below
the equilibrium temperature, resulting in a large range of
temperatures for a given grain size.  An important consideration when
dealing with this stochastically heated dust is that only some
fraction of the grains will be warm at a given time, so our mid-IR
observations may miss a substantial fraction of the dust. Grains will
be stochastically heated in this manner when two conditions are met:
1) the collisional rate is slow compared with the cooling rate of the
grain and 2) the amount of energy deposited by a single electron is
large compared with the internal energy of the grain.  

\begin{figure}
\centering
\epsscale{1.0}
\includegraphics[width=3in]{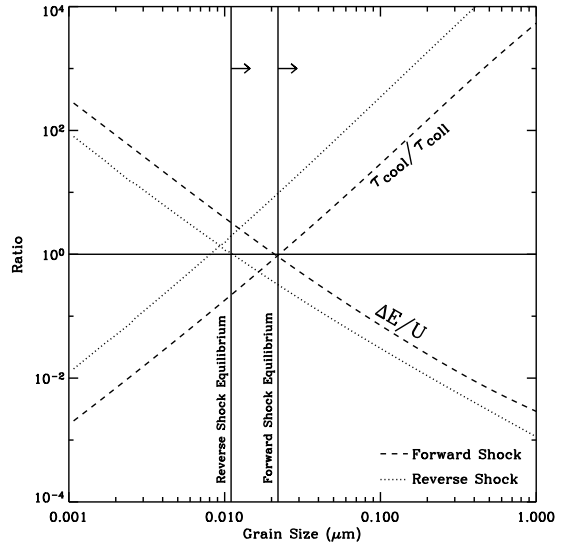}
\caption{This plot illustrates the range of dust grain sizes that will
be in equilibrium in the forward and reverse shocks.  The dotted lines
are for amorphous carbon in the reverse shock and the dashed lines are
for graphite in the forward shock, assuming the densities and
temperatures discussed in the text.  Two ratios are shown on the plot:
the ratio between the cooling time and the collision time
($\tau_{cool}/\tau_{coll}$) and the ratio between the energy deposited
by a collision with one electron and the internal energy of the grain
($\Delta E/U$).  When the ratio of the collision time to the cooling
time is large (i.e. collisions happen faster than the dust grain can
cool) and the ratio of the energy deposited by one electron to the
total internal energy of the grain is small (i.e. the grain
temperature does not change appreciably from one collision) then the
grains can be said to be in equilibrium.  The grain size for which
these conditions are met are $> 0.01$ \micron\ for amorphous carbon in
the reverse shock and $> 0.02$ \micron\ for graphite in the forward
shock.  This size is a factor of $\sim 1.5$ smaller for forsterite in
the reverse shock and $\sim 2$ smaller for silicate dust in the
forward shock.}
\label{fig:stoch}
\end{figure}

For the plasma conditions in the forward and reverse shocks,
Figure~\ref{fig:stoch} shows the ranges of grain size for which we
expect the dust to be in equilibrium with the plasma based on the
ratios between the energy deposited by one electron and the grain's
total internal energy ($\Delta E/U$) and the ratio between the cooling
time and collision time ($\tau_{cool}/\tau_{coll}$).  For simplicity
we show only one grain species for the forward and reverse shocks:
graphite in the forward shock and amorphous carbon in the reverse
shock.  The size at which equilibrium is reached is a factor of $\sim
2$ smaller for silicate grains in the forward shock and a factor of
$\sim 1.5$ smaller for forsterite grains in the reverse shock.  For
this calculation we have used the heat capacity of graphite grains
from \citet{dwek86} and silicate from \citet{draine85}, both of which
are based on fits to experimental data, and the same optical constants
used in Figure~\ref{fig:teq}.  The graphite heat capacity is used for
amorphous carbon and the silicate heat capacity is used for
forsterite.  For the estimates presented here, this level of accuracy
should be sufficient, though the heat capacity of a dust grain does
depend on its composition, shape and ``fluffiness'' among other
factors.  

The unshocked ejecta are located inside the reverse shock radius of
15''.  The density, temperature and structure of the unshocked ejecta
are unobserved.  We can place limits on the temperature of the gas and
dust by considering the processes responsible for heating and cooling
the ejecta.  In the initial phases of the remnant's expansion the
ejecta are heated by the decay of radioactive isotopes produced in the
explosion.  As the remnant expands, the ejecta cool adiabatically,
reaching a temperature cool enough to allow dust condensation after
about 400 to 800 days post-explosion \citep{todini01,nozawa03}.
Assuming adiabatic cooling with an index $\gamma = 1.25$ as used by
\citet{kozasa89} and a temperature of $\sim 1000$ K 400 days after
the supernova, the current gas temperature would be less than the
temperature of the cosmic microwave background.  Before reaching that
temperature, of course, the dust and gas would come into equilibrium
with the local radiation field which has contributions from both the
general SMC radiation field and the emission from E 0102's shock
fronts. The equilibrium temperature of big grains in the interstellar
radiation field of the SMC is $\sim 20$ K \citep{bot04,leroy07}, so we
consider this a lower limit on the temperature of the unshocked ejecta
dust.

\subsubsection{Oxygen and Neon Fine-Structure Emission
Lines}\label{sec:lines}

We observe emission from fine-structure transitions of [Ne II], [Ne III],
[Ne V] and [O IV] in the mid-IR spectrum of E 0102.  At the spectral
resolution of the LL1 order, the [O IV] line overlaps with [Fe
II] at 25.988 \micron.  However, given the predominance of oxygen in
the supernova remnant as observed in the optical, x-ray and UV and the
dearth of iron emission observed at any wavelength, we consider it
very likely that the 25.8 \micron\ emission line is due to [O IV].
Table~\ref{tab:lines} lists the detected emission lines and their
strengths.  Examination of the spatially resolved velocity information
shows a range of velocities in agreement with previous spectroscopic
measurements of optical line emission.

\begin{deluxetable*}{lccc}
\tablewidth{0pt}
\tabletypesize{\scriptsize}
\tablecolumns{4}
\tablecaption{Emission Lines in the Mid-IR Spectrum of E 0102\label{tab:lines}}
\tablehead{ \multicolumn{1}{l}{Line} &
\multicolumn{1}{l}{Measured Wavelength} &
\multicolumn{1}{l}{Central Intensity} &
\colhead{Integrated Strength} \\
\colhead{} &
\multicolumn{1}{c}{(\micron)} &
\multicolumn{1}{c}{(MJy sr$^{-1}$)} &
\multicolumn{1}{c}{($10^{-7}$ W m$^{-2}$ sr$^{-1}$)} } 
\startdata
\hbox{[Ne II]}  & 12.8053 $\pm$ 0.0005 & 6.287 $\pm$ 0.035 & 2.585 $\pm$ 0.015 \\
\hbox{[Ne III]} & 15.5514 $\pm$ 0.0010 & 2.295 $\pm$ 0.019 & 0.917 $\pm$ 0.022 \\
\hbox{[Ne V]}   & 24.3460 $\pm$ 0.0053 & 0.566 $\pm$ 0.015 & 0.234 $\pm$ 0.017 \\
\hbox{[O IV]}   & 25.9022 $\pm$ 0.0010 & 4.662 $\pm$ 0.019 & 2.132 $\pm$ 0.014 \\
\enddata
\tablecomments{These measured values are from Gaussian fits to the IRS spectra.}
\end{deluxetable*}

The contribution of [O IV] to the 24 \micron\ flux was one of the major
uncertainties for \citet{stanimirovic05} in their analysis of the mass
of newly formed dust in the remnant.  We find that [O IV] is
responsible for $\sim$ 15\% of the 24 \micron\ flux from the remnant,
significantly less than the 60\% they were forced to assume due to the
lack of spectroscopic information.

The line emission from oxygen-rich supernova remnants like E 0102 is
thought to arise from slower, radiative shocks created when the
reverse shock encounters dense clumps of ejecta. Due to the high
abundance of oxygen and other heavy elements in the nuclear-processed
stellar ejecta, the cooling time behind these shocks is shorter than
the recombination time, producing a plasma that is much cooler than
its ionization state would indicate.  The relative ionization levels
in the post-shock ejecta depend upon the details of the shock front
and the ionization state of the material prior to encountering the
reverse shock.  \citet{sutherland95} found that the optical emission
spectrum of E 0102 could be fit relatively well with a model for the
reverse shock entering a dense ejecta knot with velocities between 150
and 200 km s$^{-1}$ and creating a precursor photoionization front
that determines the initial ionization state of the pre-shock gas.
Using the observed mid-IR emission line strengths to determine
physical conditions in the ejecta would require shock modeling outside
the scope of this paper.  

\subsection{Spatial Decomposition of the Mid-IR
Emission}\label{sec:decomp}

At the resolution and signal-to-noise of our observations it is
difficult to spatially separate the emission from
circumstellar/interstellar dust in the forward shock from the emission
from newly formed dust heated by the reverse shock.  The outer radius
of the forward shock, as measured by \citet{gaetz00} is $\sim$ 22''
while the inner radius of the reverse shock is $\sim$ 15''.  In
between, the two plasmas are separated by a contact discontinuity, the
location of which has not been measured.  The resolution of the IRS
instrument in the LL1 order ranges between 6 and 9'' and the pixel
scale is $\sim$ 5''.  The angular resolution inhibits our ability to
separate newly formed dust from dust heated by the forward shock,
without more detailed modeling.  Measuring the amount of forward
shocked dust is also an interesting quantity on its own
(Section~\ref{sec:csm}).  

The close correspondence between the $0.3-10$ keV Chandra map and the
24 \micron\ dust map, as noted by \citet{stanimirovic05}, suggests
that the dust emission is mainly from the reverse shocked material and
is thus newly formed in the remnant.  In this section,  we quantify
the fraction of the dust emission at each wavelength in our cube that
can be attributed to dust in the ejecta of the supernova by
decomposing the remnant spatially using x-ray, radio and optical
observations as templates for the sources of emission in the remnant.
As discussed in the previous section, there are three mechanisms for
which we might expect to detect mid-IR dust emission: CSM/ISM dust in
the forward shocked region, newly formed dust in the x-ray emitting
regions of reverse shocked gas and newly formed dust in the dense
knots of ejecta.  This approach is similar to that used by
\citet{arendt92} in studying dust emission from the Cygnus Loop.  We
will first discuss the template images we use to represent these three
emission mechanisms.    

Emission from the forward shocked material can be traced by radio
continuum emission.  The radio continuum is synchrotron emission from
electrons interacting with the compressed CSM/ISM magnetic field in
the region between the blast wave and the contact discontinuity.
\citet{amy93} observed the remnant at 6 cm using the Australia
Telescope Compact Array.  They see a shell structure which fills the
region between the faint outer boundary of the blast wave as traced by
x-ray emission and the outer boundary of the reverse shock.   We
use the radio map as a tracer of the forward shocked circumstellar and
interstellar material.  

We have used the $0.3-10$ keV Chandra x-ray map of \citet{gaetz00} to
trace the x-ray emitting reverse shocked material.  X-ray spectroscopy
of E 0102 has demonstrated that the bright x-ray ring contains
material composed primarily of oxygen and neon with smaller
contributions from magnesium and silicon, consistent with the
nucleosynthetic products of a massive star
\citep{flanagan04,sasaki06}.  Finally, to trace the spatial
distribution of dust in dense knots of reverse shocked material we use
the optical [O III] 5007 \AA\ emission from HST ACS imaging by
\citet{finkelstein06} that arises from the aforementioned radiative
shocks. 

\begin{figure*}
\centering
\epsscale{1.0}
\includegraphics[width=5in]{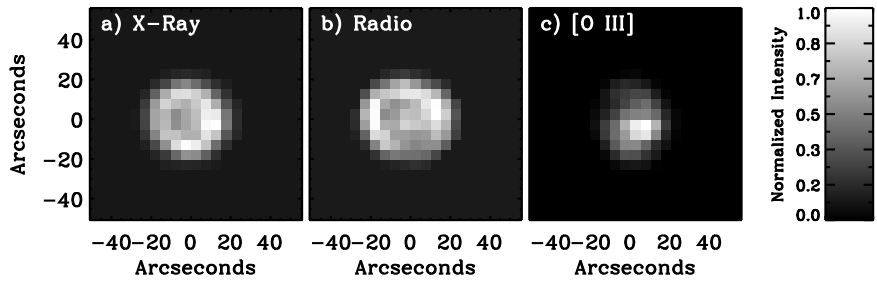}
\caption{Convolved and aligned x-ray, optical and [O III] templates used in
decomposing the mid-IR emission.  Data are from \citet{gaetz00} in the
x-ray, \cite{amy93} in the radio and \citet{finkelstein06} in the
optical.}
\label{fig:templates}
\end{figure*}

The templates were convolved to the resolution of the longest
wavelength point spread function of the IRS cube and then resampled to
the same pixel grid.  For the [O III] template a few additional
reduction steps were necessary prior to the convolving and alignment.
The processed line (F475W) and continuum (F550M) images from HST were
provided to us by S.  Finkelstein.  The wide F475W filter was used to
ensure detection of the high velocity emission from the remnant.
Relative to the point sources in the image, the remnant emission is
quite faint.  Thus, to properly construct a template it was necessary
to remove the point sources from the image by subtracting the
continuum image from the line image.  This was done by first
cross-convolving the two images with theoretical ACS PSFs from TinyTim
\citep{krist95} for the given wavebands, scaling, and subtracting.
For bright point sources, the PSF subtraction often left large
residuals, which were removed with a combination of median filtering
and masking by hand for the brightest, saturated stars.  The three
templates are shown in Figure~\ref{fig:templates} after convolution
and resampling to the LL pixel grid.  

\begin{figure*}
\centering
\epsscale{1.0}
\includegraphics[width=5in]{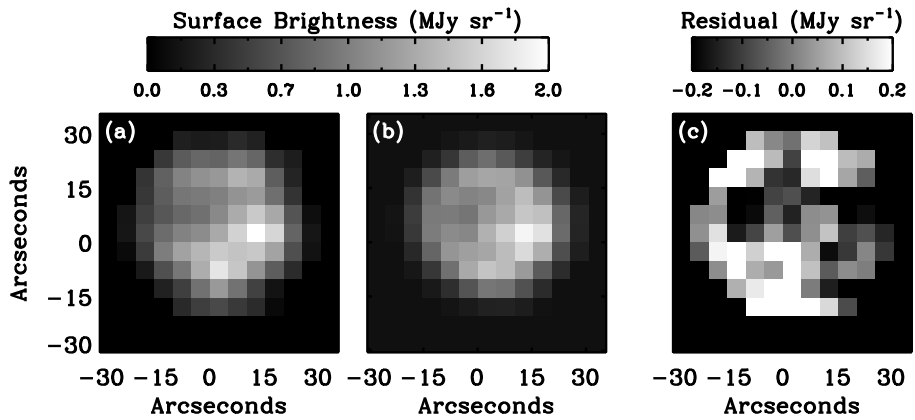}
\caption{Results of the spatial decomposition for a stack of continuum
images between 20.5 and 22.3 \micron.  We show the stacked image in
order to increase the signal-to-noise ratio for display purposes.
Panel a) shows the stack, Panel b) shows the best fit combination of
the three templates and Panel c) shows the residuals of the fit.  Note
that the residual image has a different grey scale than the images in the
Panels a) and b).  While the linear combination of the templates
reproduces most of the features of the mid-IR images, the residuals do
have some persistent features.}
\label{fig:stackcont}
\end{figure*}

\begin{figure*}
\centering
\epsscale{1.0}
\includegraphics[width=5in]{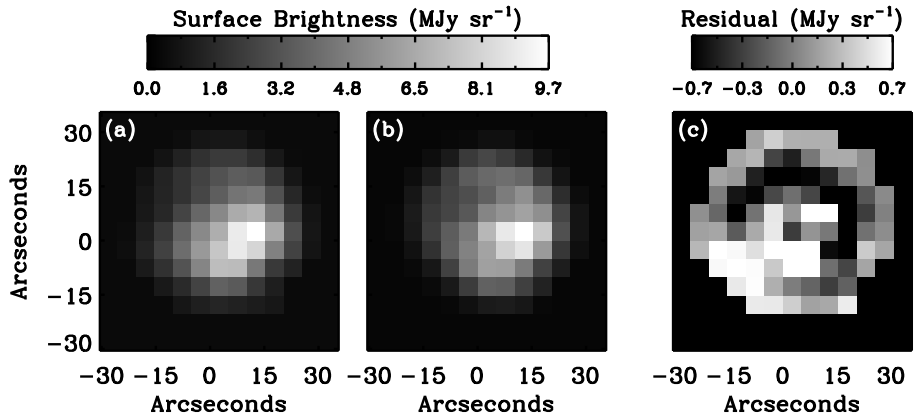}
\caption{Same as in Figure~\ref{fig:stackcont} but for the [O IV] emission
line at 25.9 \micron. Note that the grey scale for the residual image
has a different scale than the other images.  The ring of emission
visible in the residual image is most likely due to a slight
undersubtraction of the background in the vicinity of a bright
emission line.}
\label{fig:stackline}
\end{figure*}

\begin{figure*}
\centering
\epsscale{1.0}
\includegraphics[width=7in]{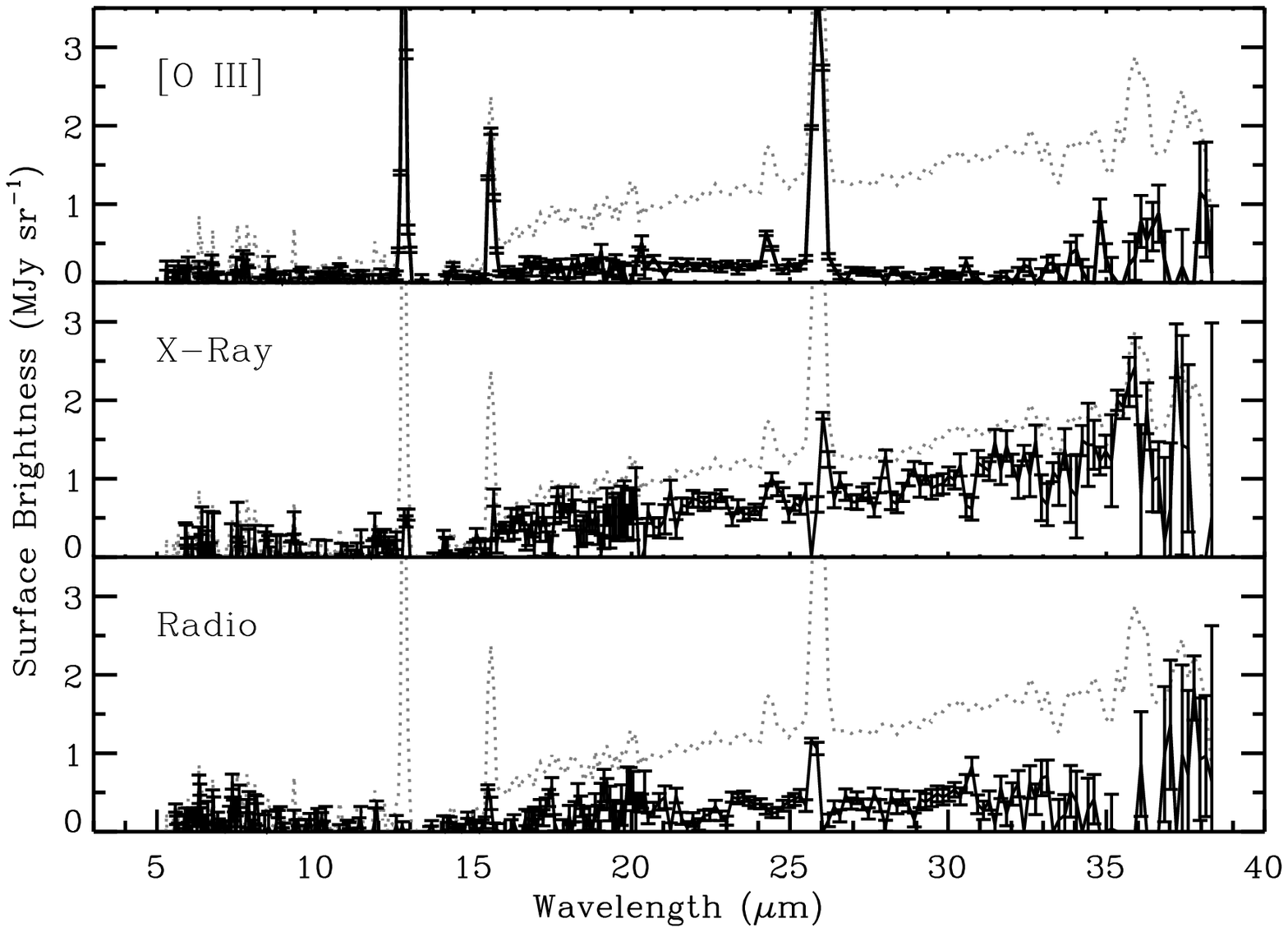}
\caption{Results of the spatial decomposition of the
mid-IR spectrum.  Each panel shows the spectrum of the emission
associated with each tracer.  Error bars include the covariance
between the tracers. Spikes in the x-ray and radio results in the
vicinity of spectral lines are artifacts of the decomposition
technique.  The dotted line overplotted in each panel shows the total
remnant spectrum.  The [O III] template traces the emission from
dense knots of reverse shocked ejecta, the x-ray template traces
shocked ejecta and the radio template traces shocked CSM/ISM.}
\label{fig:decomp}
\end{figure*}

The decomposition at each wavelength was done with a least squares fit
that constrained the coefficient of each template image to be greater
than or equal to zero.  The fit returns the best coefficients of the
linear combination of the three templates that match the observed
surface brightness distribution.  Examples of the decomposition are
shown in Figures~\ref{fig:stackcont} and~\ref{fig:stackline} for two
images which show $20.5-22.3$ \micron\ dust continuum and the [O IV]
emission line at 25.9 \micron, respectively.  To obtain the images
used for these plots we have binned the cube in wavelength to increase
the signal-to-noise of the image.  Figure~\ref{fig:decomp} shows the
results of the decomposition for all wavelengths.  The error bars
on the plot include covariance between the three templates. 
One can interpret each of these results as the spectrum of the
emission from the region described by each template: the x-ray
emitting reverse-shocked ejecta, the [O III] emitting reverse shocked
ejecta in dense knots and the forward shocked CSM/ISM material. 

It is clear from the residuals shown in Figures~\ref{fig:stackcont}
and~\ref{fig:stackline} that the correspondence between the templates
and the observations, while good, misses some features of the infrared
emission.  There is a persistently bright region in the lower left of
the images which has no counterpart in the combined template.  At the
wavelength of strong emission lines the combined template subtraction
leaves a ring of emission which most likely represents an
undersubtraction of the background at these wavelengths.  Because of
these issues, when determining the best fit dust mass we consider the
most reliable answer to be from the fit to the total remnant spectrum,
rather than the combination of the spectra associated with x-ray and
[O III] emitting ejecta.

Despite these uncertainties there are some strong conclusions that can
be drawn from the decomposition results.  One results that is a good
test for the technique is that all of the line emission is found to
come from the regions traced by optical [O III] emission, as expected.
Other results of the decomposition are: 1) that the spectrum
associated with the x-ray emitting ejecta shows dust continuum but no
emission lines, 2) there is some dust continuum in the region traced
by [O III] emission, which seems to peak at a wavelength around 22
\micron\ and 3) the forward shock spectrum also has a small dust
emission component that peaks around 20 \micron.  We discuss the
implications of the forward shocked dust further in
Section~\ref{sec:csm}.       

\subsection{Modeling the Dust Emission}\label{sec:dustmodel}

\subsubsection{Constructing the Dust Model}

Having established that the mid-IR dust continuum emission from E 0102
mainly originates in the reverse shocked ejecta of the remnant, we now
proceed to model the dust emission to attempt to place limits on the
total amount of dust formed in the CCSN.  Because we only see emission
from material that has been reheated by the reverse shock, our model
for the dust must include only the types of dust that can form out of
the nucleosynthetic materials present in the layers of ejecta that are
shocked \citep{kozasa89,nozawa03}.  A crucial observation is that the
ejecta in SNR E 0102 do not appear to be mixed.  Observations at x-ray
\citep{flanagan04,sasaki01,gaetz00,rasmussen01,hayashi94}, ultraviolet
\citep{sasaki06}, optical
\citep{finkelstein06,tuohy83,blair00,blair89} and now infrared only
detect emission from oxygen, neon, carbon, magnesium and silicon.
This is in contrast to the similar O-rich remnant Cas A, which shows
emission from sulfur, argon, calcium and iron as well
\citep{hwang04,fesen06,rho08}.  Either the nucleosynthetic layers of
the progenitor star appear to have undergone some macroscopic mixing
in Cas A or the reverse shock has encountered the oxygen-burning
layers and the carbon-burning layers, for instance, in different parts
of the supernova, simultaneously \citep{ennis06}.  The same is not
true of E 0102 where we see only nucleosynthetic products from the
carbon-burning layers.  This is most likely a consequence of the
ejecta being stratified and unmixed. 

\citet{nozawa03} model the condensation of dust in the unmixed
ejecta of core-collapse supernovae.  The types of dust they predict
are amorphous carbon in the He-rich layers; Al$_2$O$_3$, Mg$_2$SiO$_4$
(forsterite) and MgO in the O-Mg-Si layer; MgSiO$_3$ and SiO$_2$ in
the O-Si-Mg layer; and silicon and iron rich species in the deeper
nucleosynthetic layers.  To construct our model for the dust in E 0102
we first consider how deeply into the ejecta the reverse shock has
propagated, and thus what species of dust we should include in our
model.  Measurements of the abundances of oxygen, neon, magnesium and
silicon in the x-ray spectra of the shocked ejecta demonstrate that
magnesium is $\sim 2$ times more abundant than silicon
\citep{flanagan04}.  The relative amounts of these two elements
locate the reverse shock in the O-Mg-Si layer.  For this reason, in
our dust model we assume the primary species are amorphous carbon,
Al$_2$O$_3$, forsterite and MgO as predicted for the O-Mg-Si layer and
the He-rich layer by \citet{nozawa03}.   

The importance of this approach to the determination of the dust mass
in E 0102 arises from trying to find a physically motivated and yet
unique model for the dust composition.  Given enough dust species out
of the complicated mineralogy of dust condensation, one could imagine
any number of degenerate fits to the mid-IR spectrum.  By restricting
the dust model to amorphous carbon, Al$_2$O$_3$, MgO and forsterite we
hope to obtain physically meaningful values for the dust mass of each
species.  As we will discuss further below, even with these
constraints the fit is not entirely unique, demonstrating the
challenges of dealing with dust continuum emission. 

\subsubsection{Upper Limit at 70\micron}

An important constraint on our determination of the dust mass from the
mid-IR spectrum is the upper limit for the surface brightness of the
supernova remnant at 70 \micron\ from the S$^3$MC MIPS observations.
Contamination by the local foreground is higher at longer wavelengths
because of the increasing contribution from cooler dust and the lower
resolution at 70 \micron\ means it is more difficult to separate the
remnant from the shoulder of N\,76. For these reasons determining the
upper limit is not straightforward.  There is no obvious emission
visible from the supernova remnant in the MIPS 70 \micron\ image.  To
place a limit on the remnant's surface brightness we subtract a local
background using the same technique outlined in
Section~\ref{sec:reduction} with a mask region 40'' in radius to avoid
contamination from the remnant, since the angular resolution at 70
\micron\ is 18''.  The background subtraction leaves emission
consistent with zero, with an error of $\sim 1.0$ MJy sr$^{-1}$ in the
mean.  We conservatively adopt a 2-$\sigma$ upper limit of 2.0 MJy
sr$^{-1}$ at 70 \micron\ in the dust fitting to be conservative.

\subsection{Dust Model Fit Results}\label{sec:fit}

We performed a non-linear least squares fit to determine the best fit
dust model including the upper limit at 70 \micron.  We first fit to
the total spectrum of the remnant extracted over a 22'' radius region
from the background subtracted cube. Our dust model involves the four
grain species discussed above with a fixed size of 0.1 \micron.  For
dust grains in the Rayleigh limit the dust mass is independent of the
grain size, so fixing the size at 0.1 \micron\ does not affect the
fit.  The parameters of the model are the mass of dust in each species
and its temperature.  We also performed the fit on the total remnant
spectrum minus the spectrum associated with the decomposed radio
emission shown in Figure~\ref{fig:decomp}.  The results from these two
fits are very similar, as shown in Table~\ref{tab:fitresults}.
Removing the spectrum associated with the shocked CSM/ISM slightly
increases the steepness of the long wavelength continuum leading to
more cool dust in the fit.

\begin{deluxetable*}{llrc}
\tablewidth{0pt}
\tabletypesize{\scriptsize}
\tablecolumns{4}
\tablecaption{Fit Results\label{tab:fitresults}}
\tablehead{ \multicolumn{1}{l}{Spectrum} &
\multicolumn{1}{l}{Model} &
\multicolumn{1}{c}{Mass} &
\multicolumn{1}{c}{Temperature} \\
\multicolumn{1}{c}{} &
\multicolumn{1}{c}{} &
\multicolumn{1}{c}{(\msun)} &
\multicolumn{1}{c}{(K)} } 
\startdata
Total Spectrum           & Al$_2$O$_3$   & $<1 \times 10^{-6}$             & $>$ 70 \\
                         & Am. Carbon    & $\phantom{<}3 \times 10^{-3}$   & \phm{$>$}\phn 70 \\
                         & MgO           & $<1 \times 10^{-5}$             & $>$ 70 \\
	                 & Mg$_2$SiO$_4$ & $\phantom{<}2 \times 10^{-5}$   & \phm{$>$}145 \\[0.15in]
Total Spectrum           & Mg$_2$SiO$_4$ & $\phantom{<}2 \times 10^{-5}$   & \phm{$>$}145 \\
                         & Mg$_2$SiO$_4$ & $\phantom{<}1 \times 10^{-2}$   & \phm{$>$}\phn 54 \\[0.15in]
Total Spectrum $-$ CSM/ISM & Am. Carbon    & $\phantom{<}6 \times 10^{-3}$ & \phm{$>$}\phn 60 \\
                         & Mg$_2$SiO$_4$ & $\phantom{<}2 \times 10^{-5}$   & \phm{$>$}145 \\[0.15in]
X-ray Emitting           & Am. Carbon    & $\phantom{<}4 \times 10^{-3}$   & \phm{$>$}\phn 60 \\
Reverse Shocked Ejecta   & Mg$_2$SiO$_4$ & $\phantom{<}8 \times 10^{-6}$   & \phm{$>$}150 \\[0.15in]
\hbox{[O III]} Emitting  & Mg$_2$SiO$_4$ & $\phantom{<}2 \times 10^{-6}$   & \phm{$>$}180 \\
Reverse Shocked Ejecta   &               &                                 & \\[0.15in]               
CSM/ISM                  & ISM Graphite  & $\phantom{<}8 \times 10^{-6}$   & \phm{$>$}170 \\
\enddata
\tablecomments{Optical constants for amorphous carbon from
\citet{rouleau91} and ISM graphite from \citet{laor93}.  The sources
for the other optical constants can be found in
Table~\ref{tab:colddust}.}
\end{deluxetable*}

\begin{figure*}
\centering
\epsscale{1.0}
\includegraphics[width=5in]{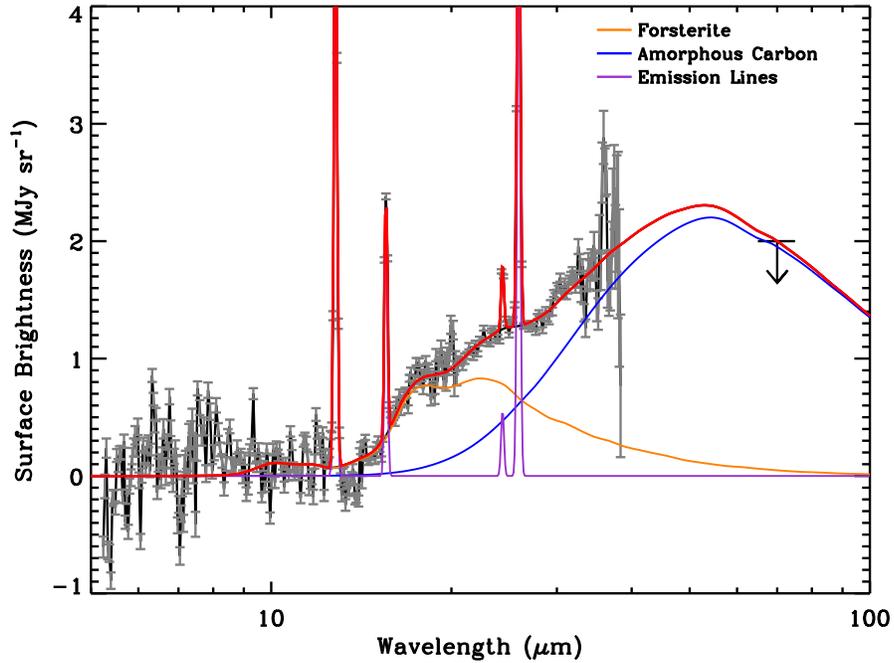}
\caption{Dust model fit to the spectrum showing $3\times 10^{-3}$
\msun\ of 70 K amorphous carbon and $2\times 10^{-5}$ \msun\ of 145 K
forsterite. The 145 K forsterite component is necessary to reproduce
the continuum shape below $\sim 30$ \micron.  The 60 K dust component
is less constrained and could be replaced with forsterite at a
comparable temperature with no impact on the quality of the fit.  The
arrow shows the upper limit at 70 \micron.}
\label{fig:allfit}
\end{figure*}

The spectrum of the remnant is well fit by a combination of forsterite
(Mg$_2$SiO$_4$) and amorphous carbon dust as shown in
Figure~\ref{fig:allfit}.  Neither Al$_2$O$_3$ nor MgO play an
important role in the best fit model.  Upper limits on the masses of
Al$_2$O$_3$ and MgO are listed in Table~\ref{tab:fitresults} at a
temperature of 70 K.  At higher temperatures the limits on the masses
are stricter.  The best fit forsterite component has a temperature of
145 K while the amorphous carbon component, which contains most of the
dust mass we detect, has a temperature of 70 K.  However, the fit
results for the cool dust component are not unique.  A fit of similar
quality using two temperature components of forsterite is listed in
Table~\ref{tab:fitresults} as well.  The continuum shape around 20
\micron\ is not well reproduced without forsterite in the dust model,
however the longer wavelength continuum shape does not distinguish
strongly between forsterite and amorphous carbon.  In the following we
proceed by assuming that the dust is amorphous carbon because 1) it
provides a more conservative estimate of the dust mass (i.e. a lower
limit) and 2) theoretical predictions suggest that there should be a
substantial amount of amorphous carbon which, if present, should all
be in the reverse shocked outer layers and visible to us in the
mid-IR.

The 70 \micron\ upper limit does not affect the amount of forsterite
we measure.  On the other hand, the limit does affect both the mass
and temperature of the amorphous carbon dust.  Decreasing the limit
tends to make the amorphous carbon dust warmer and less massive,
increasing it makes the amorphous carbon cooler and more massive.  We
note that for limits below about 1 MJy sr$^{-1}$ the fit no longer
converges properly, because the limit is inconsistent with the slope
of the data past 30 \micron.  We predict that with a small increase in
signal-to-noise and/or angular resolution the remnant should be
detectable at 70 \micron.

\begin{figure*}
\centering
\epsscale{1.0}
\includegraphics[width=5in]{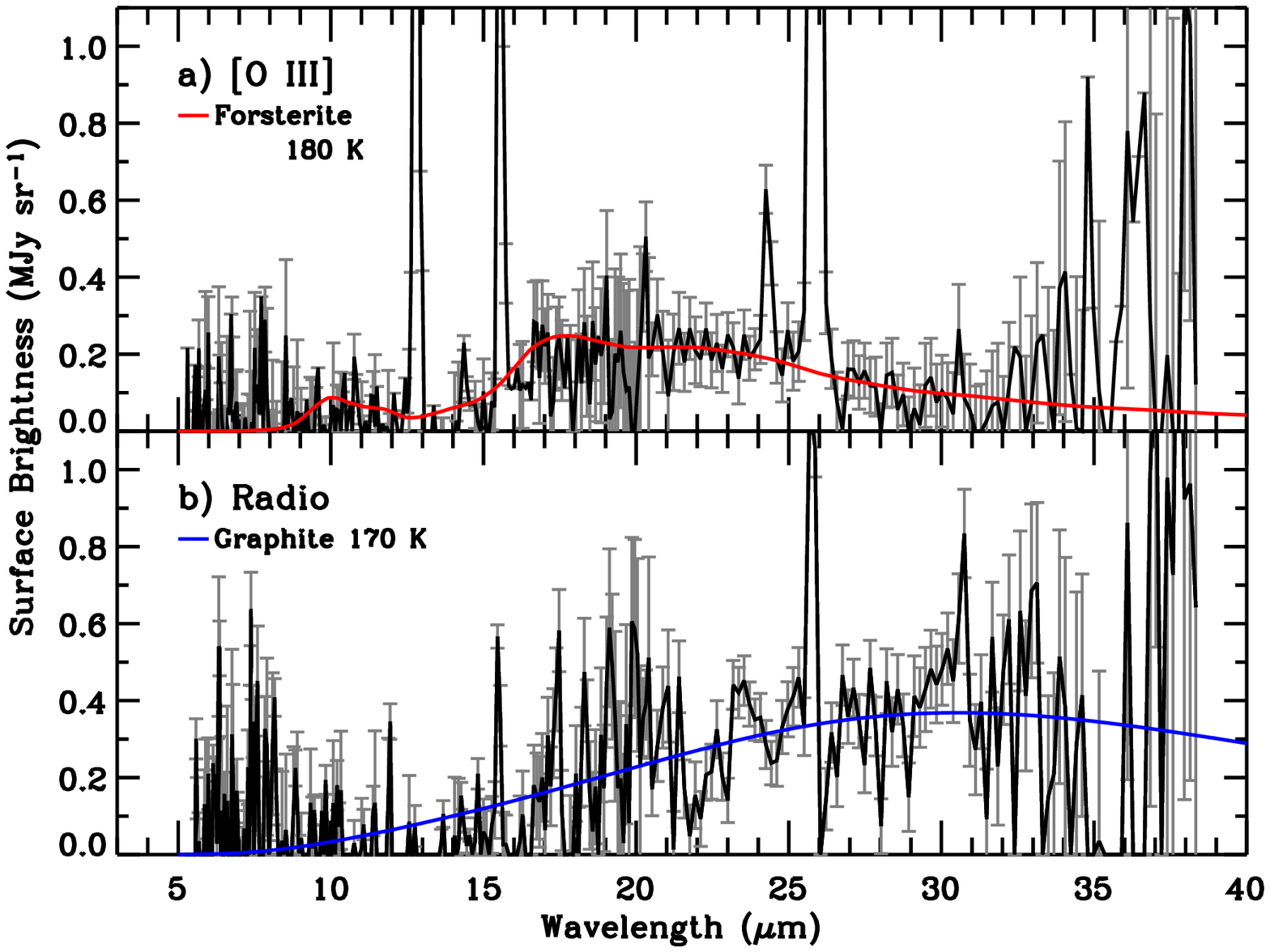}
\caption{This figure shows a zoomed in version of the spectra
associated with the [O III] and radio templates from the spatial
decomposition along with a fit to the dust continuum.  The fit to the
spectrum associated with the radio template is representative. Since
we expect it to be interstellar dust we have used interstellar
graphite from \citet{laor93}.  Silicate also reproduces the continuum
shape at the same level of quality.  For the [O III] knots, however,
the shape of the continuum is not well reproduced by amorphous carbon.
The fit using forsterite is quite good, given the uncertainties
involved.}
\label{fig:zoomed}
\end{figure*}

We also perform individual fits to the spectra associated with the
x-ray, [O III] and radio templates in order to learn about the dust in
the x-ray emitting reverse shocked ejecta, the dense knots of reverse
shocked ejecta and the CSM/ISM dust in the forward shock.
Figure~\ref{fig:zoomed} shows the dust continuum associated with the
radio and [O III] templates and the fit listed in
Table~\ref{tab:fitresults}.  The continuum shape and signal-to-noise
in the spectrum associated with the radio template do not strongly
constrain the species of dust present in the CSM/ISM material. We show
a representative fit with interstellar graphite from \citet{laor93}.
The fits imply $\sim 10^{-6} - 10^{-5}$ \msun\ of dust with a
temperature around $150-180$ K.  We discuss this component further in
Section~\ref{sec:csm}.  The dust continuum associated with the [O III]
emitting knots has a better defined shape that is well reproduced by
forsterite but not amorphous carbon.  Fits to that component yield a
dust mass of $\sim 2\times 10^{-6}$ \msun\ of forsterite at 180 K. 

\begin{figure*}
\centering
\epsscale{1.0}
\includegraphics[width=5in]{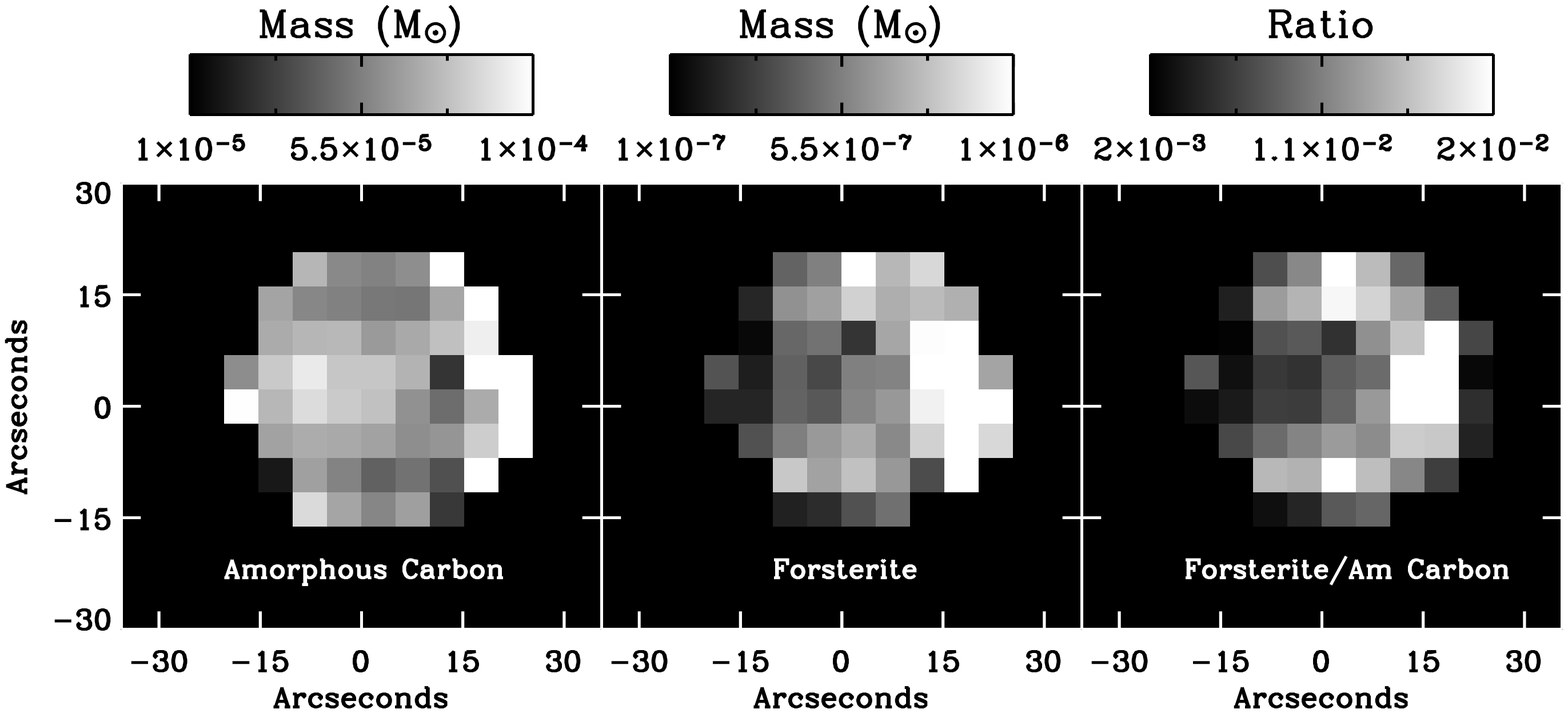}
\caption{Dust masses determined at each pixel. The first panel shows
the mass of amorphous carbon, the second shows the mass of forsterite 
and the third shows the ratio of forsterite to amorphous carbon.  The
ratio is highest at the boundary of the reverse shock which agrees
with the idea that the reverse shock is just beginning to encounter
the O-Mg-Si layer where forsterite has been produced.}
\label{fig:pbpmasses}
\end{figure*}

\begin{figure*}
\centering
\epsscale{1.0}
\includegraphics[width=5in]{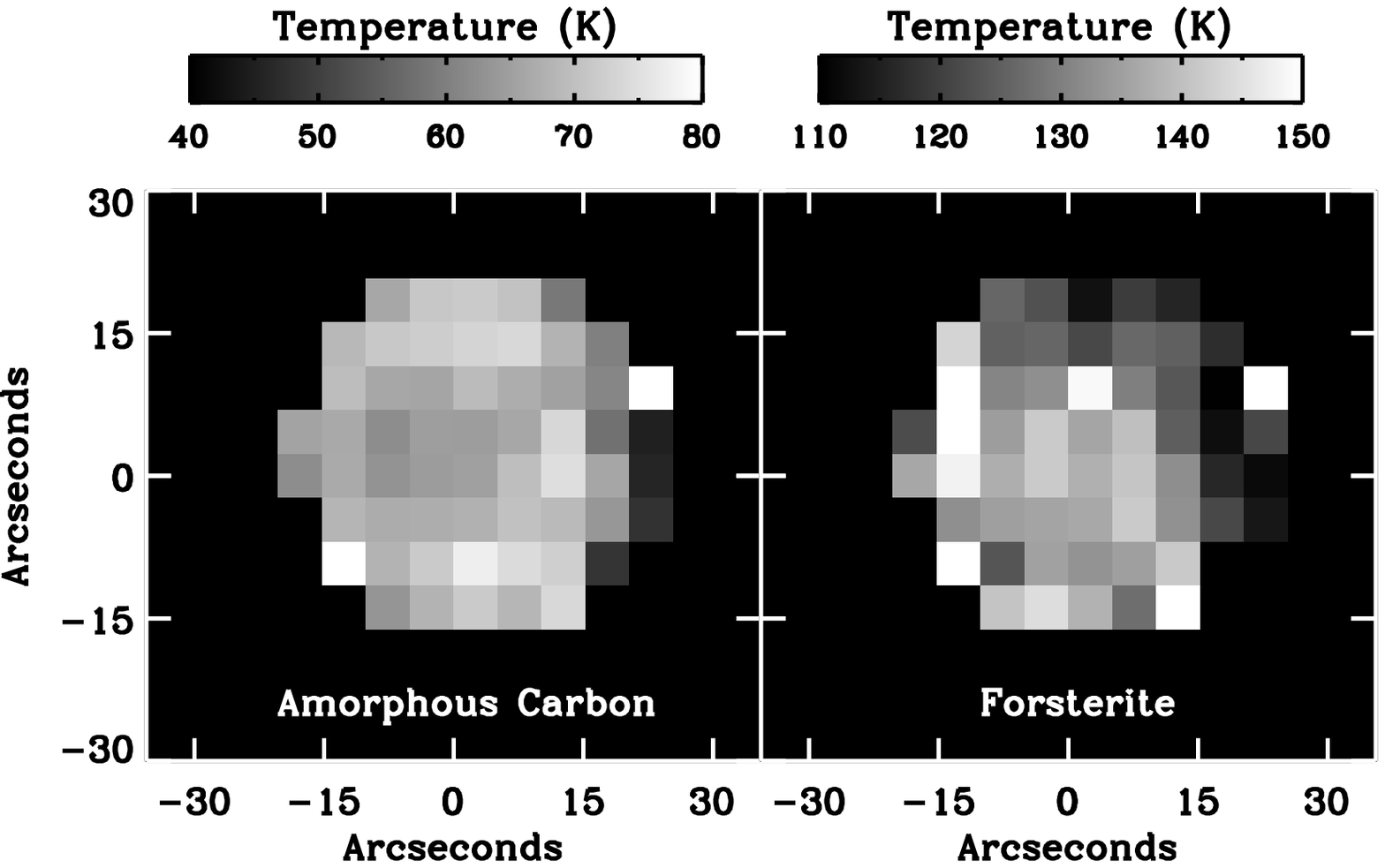}
\caption{Dust temperatures determined at each pixel.  The temperature
of the amorphous carbon dust is consistent with grains of $\sim 0.1$
\micron\ in equilibrium with the x-ray emitting reverse-shocked
plasma.  The forsterite component has temperatures which are higher
than the predicted equilibrium in the x-ray emitting plasma.  This may
be the result of the forsterite abundance increasing deeper in the
ejecta, as suggested by the forsterite to amorphous carbon abundance
shown in Figure~\ref{fig:pbpmasses}, where the recent passage of the
reverse shock has not yet destroyed smaller grains which can be
stochastically heated.}
\label{fig:pbptemps}
\end{figure*}

Using a simplified version of the model which involves only forsterite
and amorphous carbon, we performed the fit at each pixel in the map in
order to learn about the spatial distribution of the dust species.
These results are shown in Figures~\ref{fig:pbpmasses}
and~\ref{fig:pbptemps}.  The forsterite to amorphous carbon ratio
tracks the intensity of the x-ray ring of the reverse shock. This
resemblance may  be an indication that forsterite is more abundant in
the deeper layers of the ejecta which are now encountering the reverse
shock.

\section{Discussion}\label{sec:discussion}

\subsection{Newly Formed Dust}

We detect $3\times 10^{-3}$ \msun\ of 70 K amorphous carbon and
$2\times 10^{-5}$ \msun\ of 145 K forsterite in the ejecta of E 0102.
The equilibrium temperatures of grains in the x-ray emitting plasma
are between 50 and 100 K for sizes between 1.0 and 0.001
\micron.  The temperature we measure for the amorphous carbon
component corresponds to the equilibrium temperature for grains with
sizes of $\sim 0.1$ \micron. Figure~\ref{fig:stoch} verifies that
grains of this size should be in equilibrium.  

The temperature of the forsterite dust component is hotter than the
equilibrium temperature that could be achieved by grains in the x-ray
emitting plasma of the reverse shock, but is consistent with
temperatures in the dense [O III] emitting knots.  In fitting the
decomposed spectrum associated with the [O III] emission we do see
forsterite with temperatures around 180 K, typical for grains with
sizes around 0.01 \micron\ under the expected conditions.  If,
however, most of the forsterite is located in the x-ray emitting
plasma, as the results of our spatial decomposition suggest, the
temperature would imply small dust grains that are stochastically
heated.  In the following section we show that the sputtering
timescale in the reverse shocked ejecta for small grains is less than
the age of the remnant.  The presence of small, stochastically heated
dust grains in the x-ray emitting plasma can be explained if the dust
has only recently been through the reverse shock, such that it has
been in the plasma for less than the sputtering time.  Given the
results shown in Figure~\ref{fig:pbpmasses}, i.e. that the abundance
of forsterite seems to increase near the bright, inner boundary of the
reverse shock, this may very well be the explanation for why we see
forsterite grains out of equilibrium with the plasma.

\subsection{Dust Sputtering and Grain Size Constraints}\label{sec:sput}

The dust we see in E 0102 at present may not be the amount that was
initially produced in the supernova or what was initially swept up by
the blast wave.  Dust grains in shock waves can be destroyed by
sputtering, where impacts by hot gas particles erode the grain.  

For the dust in the forward shocked ISM/CSM we can estimate the
sputtering time with the relation 
\begin{equation}
\tau_{sp} \sim \frac{a}{da/dt} \sim 10^6 
\frac{a}{\micron}\left(\frac{n_e}{\mathrm{cm}^{-3}}\right)^{-1}
\mathrm{yr} , 
\end{equation}
from \citet{draine79} for gas with temperatures greater than $\sim
10^6$ K and solar abundances.  In this regime the majority of the
sputtering is done by hydrogen ions and $n_e \approx n_H$, so the
abundance of heavier elements does not significantly affect the
calculations and we can use these results for the SMC metallicity.
Following the discussion in Section~\ref{sec:temps}, we estimate
densities in the x-ray emitting forward shocked material of $\sim 1$
cm$^{-3}$.  Thus, the sputtering timescale for dust grains in the
forward shock is $\sim 10^5$ years for an 0.1 \micron\ grain and $\sim
10^3$ years for an 0.001 \micron\ grain.  Given the age of the
remnant, dust grains with sizes below 0.001 \micron\ should have been 
destroyed behind the shock.  

In the reverse shocked gas the sputtering of dust is due to collisions
with oxygen ions, since oxygen is the dominant component of the
ejecta.  We can use a simple scaling argument based on the sputtering
time relation above to estimate the lifetime of grains in the reverse
shocked material.  The sputtering rate $R$ in (g s$^{-1}$) is given by
\begin{equation} R \sim \sigma n v Y , \end{equation} where $n$ and
$v$ are the density and velocity of the sputtering particles, $\sigma$
is the grain cross section and $Y$ is the sputtering yield per
collision in units of mass per collision.  Therefore, the change of
grain radius with time, assuming constant density for the grain and a
thermal velocity for the colliding particles $v\sim (kT/m)^{1/2}$,
goes as: \begin{equation} \frac{da}{dt} \propto n T^{1/2} m^{-1/2}Y .
\end{equation} Scaling the sputtering between the forward shock and
reverse shock gas involves a number of assumptions.  First, the
density of oxygen ions in the reverse shock is not simply related to
the electron density, because the ionization state of the gas is not
in equilibrium (in which case the number of free electrons donated per
oxygen atom would be a predictable function of the temperature).
Second, the temperature of the oxygen ions is not an easily determined
quantity either, given the uncertain equilibration timescale between
the electrons and the ions after the shock.  As an estimate we can
assume $n_{\mathrm{O}} \sim n_e/3$ as a reasonable value for the
number of free electrons donated by each oxygen atom, and
$T_{\mathrm{O}} \sim T_e$, which should be the case if the
equilibration timescale between the ions and electrons is less than
$\sim 1000$ years.  In the case where the ions are hotter than the
electrons, the sputtering timescale will decrease.  \citet{nozawa07}
investigated the sputtering of dust grains in the reverse shocks of
supernova remnants and via fits to experimental data for a variety of
grain compositions and collider properties found that sputtering by
oxygen ions has $\sim 50$ times higher yield than sputtering by
hydrogen ions.  Using these values ($n_{\mathrm{O}} \sim n_e/3$, $T_e
\sim T_{\mathrm{O}}$, and $Y_{\mathrm{O}} \sim 50 Y_{\mathrm{H}}$) we
estimate that the sputtering timescale in the reverse shocked ejecta,
which has an electron density of $\sim 20$ cm$^{-3}$, for a grain of
size $0.1$ \micron\ is $\sim 1000$ years. Thus, in the x-ray emitting,
oxygen-rich ejecta, the destruction time for grains smaller than 0.1
\micron\ is shorter than the age of the remnant and we expect that the
grain size distribution in the plasma has been altered significantly.  

The amorphous carbon dust, which makes up the majority of the mass we
detect from the reverse shocked ejecta, has a temperature which
suggests emission from $\sim 0.1$ \micron\ grains.  Based on our
estimate of the sputtering timescale in the reverse shocked plasma,
these grains would be able to survive the lifetime of the remnant.
The warmer forsterite dust component is found to be associated 
with the dense [O III] knots, which have a higher equilibrium
temperature due to their high densities, and with the most recently
reverse-shocked ejecta, in which smaller dust grains may still exist
and be stochastically heated to higher temperatures.  

\subsection{The Possibility of Cold Dust}

\begin{figure*}
\centering
\epsscale{1.0}
\includegraphics[width=5in]{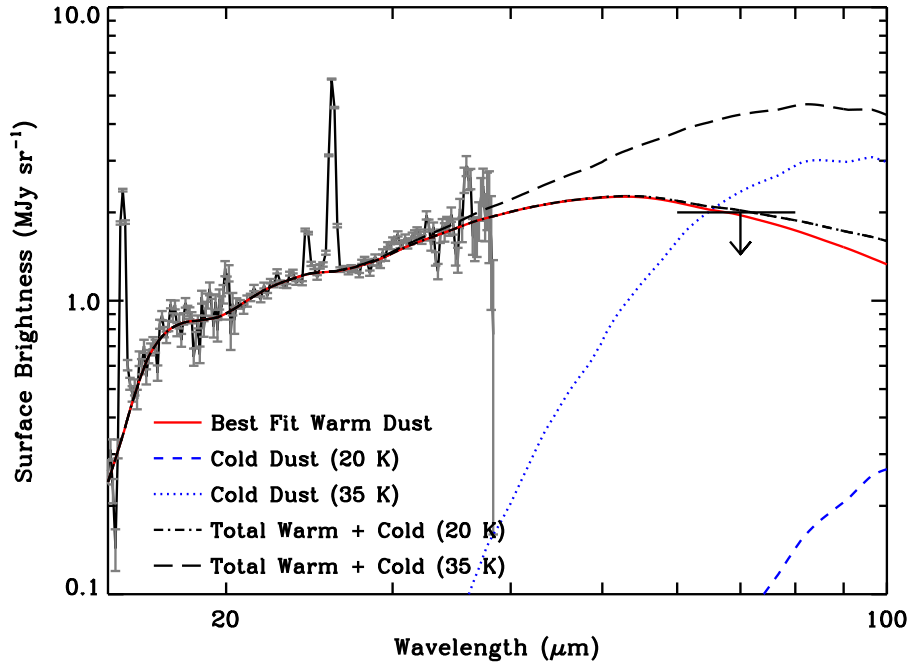}
\caption{This figure shows predictions for the far-IR spectrum of dust
in the remnant using the cold dust model components listed in
Table~\ref{tab:colddust}.  The cold dust spectrum contains the total
dust mass predicted by the \citet{nozawa03} models for a 25 \msun\
progenitor excluding amorphous carbon.  The dashed line shows this
model assuming the cold dust has a temperature of 20 K and the
dotted line shows the same model with a temperature of 35 K.  The 
black dash-dot and long-dashed lines show the total spectrum of cold
dust plus the warm dust we detect in the mid-IR.  This plot
demonstrates that all of the predicted $\sim 0.6$ \msun\ of dust can
be present in the remnant if it has a temperature of 20 K, showing
that our current observations alone can not strongly constrain the
dust condensation efficiency.}
\label{fig:colddust}
\end{figure*}

Because our observations are only sensitive to warm, recently shocked
dust, we cannot quantify the total amount of newly formed dust
produced in the remnant.  To understand the limits that the 70
\micron\ observation places on the mass of cold dust, we construct a
model for the unshocked dust in the supernova remnant based on the
condensation models of \citet{nozawa03} for a 25 \msun\ progenitor.
We list the dust species included in the model, the source for their
optical constants, and the masses predicted by \citet{nozawa03} in
Table~\ref{tab:colddust}.  The total mass of dust predicted by their
models is 0.6 \msun, minus the amorphous carbon component which we
argue must primarily be in the outer, shocked layers of the ejecta.
In Figure~\ref{fig:colddust} we show the remnant's spectrum, the best
fit model with amorphous carbon and forsterite from
Section~\ref{sec:fit}, and the spectrum of the cold dust at
temperatures of 20 (blue) and 35 K (black). It is clear that our limit
at 70 \micron\ does not preclude the possibility of a substantial mass
of cold dust in the remnant, since at a temperature of 20 K, the
entire predicted mass of dust can be hidden at 70 \micron.
Determining the total amount of newly formed dust in E 0102 will
require longer wavelength observations. 

\begin{deluxetable*}{lccc}
\tablewidth{0pt}
\tabletypesize{\scriptsize}
\tablecolumns{4}
\tablecaption{Cold Dust Model Components\label{tab:colddust}}
\tablehead{ \multicolumn{1}{l}{Species} &
\multicolumn{1}{l}{Mass} &
\multicolumn{1}{l}{Optical Constants Reference} \\
\multicolumn{1}{c}{} &
\multicolumn{1}{c}{(\msun)} &
\multicolumn{1}{c}{} } 
\startdata
Mg$_2$SiO$_4$ & 0.2 & \citet{jaeger03} \\
Amorphous Silicon & 0.2 & \citet{piller85} \\
SiO$_2$ & 0.07 & \citet{philipp85} \\
FeS & 0.07 & \citet{begemann94} \\
MgO & 0.06 & \citet{roessler91} \\
Fe & 0.02 & \citet{lynch91} \\
MgSiO$_3$ & 0.003 & \citet{dorschner95} \\
Al$_2$O$_3$ & 0.002 & \citet{begemann97} \\
\enddata
\end{deluxetable*}

\subsection{Implications for Dust Production in CCSN and Comparison
with Previous Results}

\citet{stanimirovic05} found a mass of $8\times 10^{-4}$ \msun\ of
dust in E 0102 based on the flux at 24 \micron.  Using upper limits at
8 and 70 \micron, they argued that the dust had to have a temperature
around 120 K.  They assumed that 60\%  of the 24 \micron\ emission was
from the contribution of [O IV], a value we measure to be 15\%.  These
differences alone do not explain the $\sim 4$ times more dust that we
find compared to \citet{stanimirovic05}.  Instead, the greater
contribution of  cooler dust with a temperature of around 70 K
accounts for most of the difference.  Our measurement of the
temperature and mass of forsterite dust are comparable to their
predictions.  The spectrum past 30 \micron, however, is not compatible
with only a $\sim 120$ K component.   

Similar mid-infrared spectral mapping observations have been carried
out on Cas A, another young, oxygen-rich CCSN remnant, by
\citet{rho08}.  There are some important differences between the
observations.  Cas A's relative proximity and brightness allows more
detailed fits to the spectrum of the remnant. Cas A also has emission
from S, Si, and Ar in its spectrum  in addition to O, Ne, and Mg.
This could be the result of macroscopic mixing, as suggested by
\citet{douvion99}, or of velocity inhomogeneity (Smith et al., in
prep.), causing different layers of the ejecta to encounter the
reverse shock simultaneously.  In both of these situations, the
overall composition of the nucleosynthetic layers is preserved and the
same species of dust should have formed in the ejecta as in E 0102. It
may not be possible, however, to separate different nucleosynthetic
layers spatially, making the modeling more complex.  

\citet{rho08} find on the order of $0.02-0.05$ \msun\ of dust in the
Cas A remnant, with contributions from a variety of dust species.
They find three distinct spectra in Cas A, a 21 \micron\ peak spectrum,
a weak 21 \micron\ spectrum and a featureless spectrum.  The weak 21
\micron\ spectrum has associated strong neon lines, similar to what we
see in E\,0102.  The dominant dust components by mass that
\citet{rho08} fit to the weak 21 \micron\ spectrum are two temperature
components of amorphous carbon at $\sim 80$ and $\sim 200$ K totaling
$\sim 1-2\times 10^{-3}$ \msun\ and $\sim 0.6-1.4\times 10^{-2}$
\msun\ of FeO at $\sim 60$ K.  The mass of amorphous carbon we find is
comparable to their results, however we have no evidence of FeO in our
spectrum and it seems unlikely that a substantial amount of FeO could
be produced, since it requires microscopic mixing of the ejecta.
Comparing with the total amount of dust that \citet{rho08} find in Cas
A, we see  similar amounts of both amorphous carbon and Mg$_2$SiO$_4$,
on the order of $10^{-3}$ and $10^{-5}$ \msun\ respectively.  However,
they argue for significant amounts of Fe, FeO, FeS and SiO$_2$,
producing a total dust mass of between $0.02-0.05$ \msun\ of
newly-formed dust---an order-of-magnitude larger than what we observe
in E\,0102.  The progress of the reverse shock into the deeper
nucleosynthetic layers of Cas A may partially account for the
difference in the observed dust mass between the two remnants.  It is
possible that similar dust masses could be present but the iron- and
silicon-rich dust species are in unshocked layers of the ejecta of
E\,0102 and are too cold for us to detect in these observations.

Given the progress of the reverse shock into the ejecta, we can claim
with good confidence that we should see essentially all of the
amorphous carbon dust that has been produced in the mid-IR.  This
being the case, we find substantially less amorphous carbon dust than
predicted by dust condensation models.  The models of \citet{nozawa03}
predict on the order of $10^{-1}-10^{-2}$ \msun\ of amorphous carbon,
whereas we only see $\sim 3\times 10^{-3}$ \msun.  We estimate that
the sputtering time for dust in the reverse-shocked ejecta is less
than the age of the remnant, so we may be observing less dust then was
initially formed.  \citet{nozawa07} calculate the dust destruction via
sputtering in the reverse shock for the same models quoted above using
a simple prescription for the interaction of the ejecta with the
surrounding ISM.  They find that for an ambient density of $0.1$
cm$^{-3}$, only 8\% of the amorphous carbon dust is destroyed for
their 25 \msun\ progenitor.  Most of the mass of carbon dust in that
model is in grains with sizes greater than 0.05 \micron, which have
longer sputtering lifetimes.  We have no constraints on the initial
grain size, so it is difficult to estimate how much of the dust in the
remnant has been destroyed up to this point.  

Previous observations of newly formed dust in CCSN have yielded dust
masses in the range of $10^{-3}$ to $10^{-5}$ \msun, with some
controversial measurements claiming higher masses.  The majority of
the measurements come from CCSN of Type II or their remnants, which
have hydrogen envelopes and red supergiant progenitors.  Dust
formation has been observed in two Type Ib events: SN 1990I
\citep{elmhamdi04} and the peculiar event SN 2006jc
\citep{smith08,nozawa08} which shows evidence for dust formation at
very early times, around 50 days after the explosion.  The
masses determined from observations of 2006jc range from a few
$10^{-6}$ to a few $10^{-3}$ depending on the observation date,
wavelength and dust model.  Recently, \citet{nozawa08} have modeled
dust condensation in a Type Ib event in order to assess the effects of
a small or non-existent hydrogen envelope on the dust condensation
efficiency.  The high expansion velocities mean that the density drops
quickly, leading to dust condensation at earlier times in these
events.  The condensation efficiency is high, however, yielding $0.7$
\msun\ of amorphous carbon in their model of SN 2006jc, which is in
contrast to the relatively small amount of dust detected in the
near-IR and mid-IR observations.  Our results for E 0102 lie near the
upper bounds of the observed dust mass in SN 2006jc. It may not be the
case, however, that dust is forming in the ejecta of SN 2006jc.
\citet{smith08} argue that the dust condensation occurs in the
cooling region of the shocks created by the interaction of the
blast wave with an expanding shell of circumstellar material. If this
is the case, we may not expect to see comparable species or amounts of
dust in SN 2006jc and E 0102.

\subsection{The Circumstellar Environment of SNR E
0102}\label{sec:csm}

The progenitor of SNR E 0102 was most likely a massive star that
lost most of its hydrogen/helium envelope before exploding as a Type
Ib/Ic or IIb/L supernova \citep{blair00,chevalier05}.  As such, it
would have significantly impacted the circumstellar medium in the
form of winds and mass loss.  Here we briefly discuss what we can
learn about the CSM/ISM from our observations.

The spectrum associated with the radio continuum template, as
discussed in Section~\ref{sec:decomp}, shows the dust continuum
emission from forward shocked CSM/ISM material.  Fits to this spectrum
with interstellar graphite and silicate grains give masses in the
range $10^{-5}-10^{-6}$ \msun\ and temperatures of $\sim 180$ K.  If
the remnant were expanding into the average SMC interstellar medium
with a density of $n \sim 0.1$ cm$^{-3}$, we would expect
approximately 2 \msun\ of swept up gas and $\sim 3\times 10^{-3}$
\msun\ of swept up dust at this point in time \citep[using a dust to
gas ratio of 1/700 as determined by][]{leroy07}.  Although the fit
dust mass is far less than this expected value, the temperature
indicates that the dust we see is not in collisional equilibrium and
therefore we are only detecting the fraction of the dust that is warm 
at a given time (see Figure~\ref{fig:teq} for the predicted
equilibrium temperature in the forward shock).  

For stochastic heating, the fraction of grains which are warm at any
time is given approximately by the ratio of the collision time to the
cooling time $f \sim \tau_{cool}/\tau_{coll}$.  This ratio is shown as
a function of grain size in Figure~\ref{fig:stoch}.  Grains with $a >
0.02$ \micron\ will not be stochastically heated and will have
equilibrium temperatures between 40 and 70 K.  In
Section~\ref{sec:sput}, we estimated that grains with $a < 0.001$
\micron\ should be destroyed by sputtering on timescales shorter than
the age of the remnant.  Within these limits $0.001 < a < 0.02$, we
can estimate that $f \sim 0.01-1$, i.e. that we detect between 1 and
100\% of the dust mass in the forward shocked CSM/ISM. Thus, the
amount of dust in the forward shocked CSM/ISM is essentially
unconstrained by our observations, ranging from $10^{-6} - 10^{-3}$
\msun. 

Depending on the progenitor star and its winds, we may not expect to
see any dust in the forward-shocked plasma.  This would be the case if
the blast wave has not yet reached the pristine ISM and the
circumstellar medium has been cleared of dust by the winds of the
progenitor star.  \citet{chevalier05} argue that E\,0102 is still
interacting with the wind of its progenitor star based on the
positions of the forward and reverse shocks and the lack of limb
brightening in the x-ray emission from the forward shock.  The dust
content of the circumstellar medium will depend on the duration of the
wind and the previous evolution of the progenitor (for example,
whether there were outbursts similar to those observed from the
progenitor of SN 2006jc \citep{pastorello07}) among other factors.
Further observational constraints on the progenitor mass and evolution
will be necessary to understand whether the dust we see associated
with the forward shock fits with the overall picture of the evolution
of E\,0102.

\section{Summary \& Conclusions}\label{sec:conclusions}

To summarize, we present mid-IR spectral mapping observations of the
supernova remnant 1E\,0102.2$-$7219 in the vicinity of N\,76 in the
Small Magellanic Cloud.  The $\sim 1000$ year old, oxygen-rich remnant
is thought to be the product of a Type Ib/Ic or IIb/L supernova.
Currently the reverse shock, produced by the dynamical interaction of
the ejecta with the surrounding ISM/CSM, is propagating back through
the ejecta.  Based on spectroscopic observations at x-ray,
ultraviolet, optical and now infrared wavelengths, the ejecta down to
the depth of the reverse shock appear to be composed of oxygen, neon,
magnesium, silicon and carbon.  This composition indicates that the
reverse shock has only penetrated into the deeper layers of
nucleosynthetic products and that no large-scale mixing of deeper
layers has occurred.

The mid-IR spectrum of the remnant shows fine-structure emission lines of
oxygen and neon on top of emission from warm dust.  We use observations
of the remnant at x-ray, radio and optical wavelengths as templates to
decompose the infrared emission into components associated with the
newly-formed dust in the reverse shocked ejecta and ISM/CSM dust in
the forward shocked region.  Our decomposition of the spectrum shows
(1) dust continuum associated with the x-ray emitting reverse shocked
ejecta, (2) oxygen and neon fine-structure emission lines and warm dust
continuum associated with the optical [O III] emitting gas in dense
knots of ejecta where the reverse shock is slow and radiative and (3)
a small amount of dust emission associated with the forward shock.
The majority of the dust is associated with the x-ray emitting reverse
shocked material.

We construct a model for the newly-formed dust in the unmixed ejecta.
Using x-ray measurements of the abundance of magnesium and silicon we
estimate the location of the reverse shock to be in the O-Mg-Si layer.
We model the dust emission with amorphous carbon, forsterite
(Mg$_2$SiO$_4$), Al$_2$O$_3$, and MgO as appropriate for the
composition of the ejecta down to the depth of the O-Mg-Si layer,
motivated by the results of dust condensation models.  The best fit
model contains $3\times 10^{-3}$ \msun\ of amorphous carbon at a
temperature of 70 K and $2\times 10^{-5}$ \msun\ of forsterite with
temperature of 145 K.  The masses of Al$_2$O$_3$ and MgO are limited
to be less than $1\times 10^{-6}$ and $1\times 10^{-5}$ \msun\ at a
temperature of 70 K, respectively, since they do not contribute
substantially to the continuum shape in the mid-IR.  The temperature
of amorphous carbon agrees with expectations for 0.1 \micron\ grains
in the x-ray emitting plasma behind the reverse shock, which agrees
with our estimate  that smaller grains have been destroyed by
sputtering.  The temperature of the forsterite component suggests that
the emission is from smaller, stochastically heated grains which have
recently been heated by the passage of the reverse shock or are
residing in dense knots of ejecta material.

If the reverse shock is presently encountering material from the
O-Mg-Si layer of the progenitor we should be detecting the majority of
the newly-formed amorphous carbon dust, since it forms primarily in
the outermost layers of the ejecta.  Comparison with the
\citet{nozawa03} results show substantially less amorphous carbon dust
in E 0102 than would be predicted based on the progenitor mass.  Our
results for the masses of amorphous carbon and forsterite are
comparable to work on Cas A by \citet{rho08}.

Our mid-IR observations are not sensitive to cold dust present in the
remnant.  We show that if the dust has a temperature of 20 K, $\sim
0.6$ \msun\ of dust could be present in the remnant and undetectable
in our observations.  Observations at longer wavelengths by Herschel,
LABOCA on APEX or ALMA will be necessary to determine the amount of
cold dust present in the remnant.  Without a substantial mass of
hidden cold dust, the dust production in E\,0102 falls orders of
magnitude below what would be necessary to explain the observations of
dust around high redshift quasars.

\acknowledgements

The authors would like to thank S. Finkelstein for providing us with
the [O III] map.  K.S. would like to thank Eli Dwek, Maryam Modjaz,
and Nathan Smith for useful conversations and Adam Leroy for providing
a quiet place to work at a crucial time.  K.S.  acknowledges the
support of an NSF Graduate Research Fellowship.  This work is based on
observations made with the Spitzer Space Telescope, which is operated
by the Jet Propulsion Laboratory, California Institute of Technology
under a contract with NASA. This research was supported in part by by
NASA through awards issued by JPL/Caltech (NASA-JPL Spitzer grant
1264151 awarded to Cycle 1 project 3316, and grants 1287693 and
1289519 awarded to Cycle 3 project 30491).

{\it Facilities:} \facility{Spitzer ()}



\end{document}